\documentclass[preprint,review,12pt]{elsarticle}
\usepackage[left=2cm,right=2cm, top=2.7cm, bottom=2.7cm]{geometry}

\biboptions{sort&compress}
\usepackage{graphicx}
\usepackage{color}
\usepackage{floatrow}
\usepackage{subfig}
\usepackage{mwe}
\usepackage{caption}
\usepackage{afterpage}
\usepackage{changepage}
\usepackage[hyphens]{url}
\usepackage{hyperref}
\hypersetup{
     colorlinks   = true,
     citecolor    = blue
}
\usepackage[version=4]{mhchem}

\usepackage{ulem}

\usepackage{amssymb}
\usepackage{amsthm}
\usepackage{amsmath,bm}
\allowdisplaybreaks[4]
\usepackage{bibentry}
\usepackage{mathrsfs}

\usepackage[labelsep=period,labelfont=bf,figurename=Fig.]{caption}
\usepackage{tikz}
\usepackage{pgfplots}
\usetikzlibrary{arrows,shapes,trees}
\usetikzlibrary{decorations.pathreplacing}
\usetikzlibrary{intersections,backgrounds}
\usetikzlibrary{calc}
\usetikzlibrary{positioning}
\usepackage{tkz-euclide}
\tikzset{>=latex}
\usepackage{colortbl}

\usepackage{pict2e}
\usepackage{overpic}
\usepackage{multirow}
\usepackage{calc}
\usepackage{longtable}
\usepackage{pbox}
\usepackage{array}
\newcolumntype{L}[1]{>{\raggedright\let\newline\\\arraybackslash\hspace{0pt}}m{#1}}
\newcolumntype{C}[1]{>{\centering\let\newline\\\arraybackslash\hspace{0pt}}m{#1}}
\newcolumntype{R}[1]{>{\raggedleft\let\newline\\\arraybackslash\hspace{0pt}}m{#1}}

\usepackage{etoolbox}
\usetikzlibrary{arrows}

\makeatletter
\def\ifGm@preamble#1{\@firstofone}
\appto\restoregeometry{%
  \pdfpagewidth=\paperwidth
  \pdfpageheight=\paperheight}
\apptocmd\newgeometry{%
  \pdfpagewidth=\paperwidth
  \pdfpageheight=\paperheight}{}{}
\makeatother

\usepackage[flushleft]{threeparttable}
\usepackage{booktabs}
\usepackage{longtable}

\newlength{\bibitemsep}
\newlength{\bibparskip}
\let\oldthebibliography\thebibliography
\renewcommand\thebibliography[1]{%
  \oldthebibliography{#1}%
  \setlength{\parskip}{\bibitemsep}%
  \setlength{\itemsep}{\bibparskip}%
}

\tikzstyle{cloud} = [draw, circle, fill=blue!20]
\tikzstyle{cloud1} = [draw, circle, fill=red!20]

\journal{Combustion and Flame}

\begin{document}

\begin{frontmatter}



\title{Thermodiffusively unstable laminar hydrogen flame in a sufficiently large 3D computational domain --- Part I: Characteristic patterns}


\author[itv]{Xu Wen\corref{cor1}}
\ead{x.wen@itv.rwth-aachen.de}
\author[itv]{Lukas Berger}
\author[tju]{Liming Cai}
\author[ulb1,ulb2]{Alessandro Parente}
\author[itv]{Heinz Pitsch}

\address[itv]{Institute for Combustion Technology, RWTH Aachen University, Aachen 52056, Germany}
\address[tju]{School of Automotive Studies, Tongji University, Cao'an Road 4800, 201804 Shanghai, China}
\address[ulb1]{Aero-Thermo-Mechanics Laboratory, Universit\`e Libre de Bruxelles, Brussels 1000, Belgium}
\address[ulb2]{Brussels Institute for Thermal-Fluid Systems and Clean Energy, Universit\`e Libre de Bruxelles, Brussels 1000, Belgium}
\cortext[cor1]{Corresponding author:}

\begin{abstract}
Thermodiffusive instabilities can have a leading order effect on flame propagation for lean premixed hydrogen flames. Many simulation studies have been performed to study this effect, but almost exclusively in two-dimensional (2D) or domain sizes too small to support the characteristic large-scale features of the instability. The main purpose of this study is to quantify the differences of 3D and 2D flames using simulations on sufficiently large domains. To this end, direct numerical simulations (DNS) of thermodiffusively unstable laminar premixed hydrogen flames stabilized in 3D domains were performed. The effects of confinement on the flame dynamics are rigorously investigated by varying the domain size. The flame burning velocity shows a strong dependence on the domain size when the lateral domain width is less than 50 laminar flame thicknesses. The characteristic patterns of the thermodiffusively unstable flame are analyzed in detail, including the instantaneous flame structure, global burning velocity, flame surface area, stretch factor, curvature distribution, and the cell size. The effects of the computational setup (2D vs.~3D) on the local flame front curvature and the distributions of the thermo-chemical quantities are quantified through a conditional analysis. The formation and destruction mechanisms of the distinct cellular structures observed in the 3D domain are analyzed focusing on the interactions between the flame dynamics and the flow field, and the contributions of the production of flame surface density, kinematic restoration, and curvature dissipation are quantified. Compared with the corresponding 2D simulation, the flame surface area in the cut plane of the 3D configuration is similar, yet the burning velocity and the stretch factor are increased. The reason for this is the difference in curvature statistics in 2D and 3D. In 3D, larger extreme curvature values are obtained, leading to quantitatively different distributions of the thermo-chemical variables. For example, the peak H radical concentration in the 3D simulation is about twice higher than in the 2D simulation. 
\end{abstract}


\end{frontmatter}


\section*{Novelty and significance statement}

The novelty of the present work includes the following aspects:
\begin{enumerate}[(\romannumeral1)]
\item First large-scale 3D DNS of laminar lean premixed hydrogen flame on domain large enough to obtain domain-independent results;
\item First quantitative assessment of characteristic patterns of 3D thermodiffusively unstable premixed hydrogen flames;
\item First quantitative assessment of differences between 2D and 3D thermodiffusively unstable premixed hydrogen flames for domain-independent results.
\end{enumerate}

\section*{Author contributions}

X.~Wen performed simulations, analyzed data, and wrote the paper. L.~Berger assisted with the simulations, data analysis, reviewing, and editing. L.~Cai implemented chemical mechanism reduction. A.~Parente contributed to the reviewing, editing, and supervision. H.~Pitsch contributed to the data analysis, reviewing, editing, and supervision.

\section{Introduction}
\label{Sec:1}

The usage of hydrogen offers a solution to carbon-free industrial processes, supporting the aim for an economy with net-zero greenhouse gas emissions. Thermo-chemical conversion is a promising way to harvest the chemical energy of the hydrogen molecule in industrial applications. Compared with non-premixed or stoichiometric combustion, fuel-lean premixed combustion generates less NO$_\mathrm{x}$ emissions \citep{correa1993review}. Challenges in lean hydrogen combustion predominately arise from the augmented effects of preferential diffusion and the induced intrinsic instabilities, which can substantially change flame dynamics and heat release rates. The intrinsic instabilities include two mechanisms \citep{matalon2007intrinsic, law2000structure}, i.e., hydrodynamic instability due to the density jump across the flame front and thermodiffusive instability due to the disparity between heat flux and mass flux entering or leaving the reaction zone. The thermodiffusive instability is strong only in fuel-lean hydrogen flames where the effective Lewis number of the unburnt mixture is low. 

Various DNS studies have been conducted for thermodiffusively unstable premixed hydrogen flames focusing on different aspects \citep{haworth1992numerical, baum1994direct, trouve1994evolution, chen2000stretch, im2002preferential, tanahashi2002local, kadowaki1996numerical, kadowaki1999lateral,  kadowaki2002formation, kadowaki2005numerical, kadowaki2005unstable, bell2007numerical, day2009turbulence, day2011numerical, aspden2011characterization, altantzis2012hydrodynamic, altantzis2013numerical, frouzakis2015numerical, uranakara2016flame, yu2017nonlinear, zhou2017effect, kadowaki2021effects, berger2019characteristic, berger2022intrinsicPart1, berger2022intrinsic, wen2022flamepart1, wen2022flamepart2,  you2020modelling, berger2022synergistic, rieth2023effect, dinesh2016high, song2022diffusive, lee2022dnsPart1, lee2022dns, howarth2022empirical, howarth2023thermodiffusively, lu2021modeling}. Haworth and Poinsot \citep{haworth1992numerical}, Baum et al.~\citep{baum1994direct}, Trouv{\'e} and Poinsot \citep{trouve1994evolution}, Chen and Im \citep{chen2000stretch}, Im and Chen \citep{im2002preferential}, and Tanahashi et al.~\citep{tanahashi2002local} conducted pioneering works for thermodiffusively unstable premixed hydrogen flames. Kadowaki and co-authors \citep{kadowaki1996numerical, kadowaki1999lateral, kadowaki2002formation, kadowaki2005unstable} conducted a series of earlier works to investigate the unstable behaviors of cellular premixed flames induced by intrinsic instability in both 2D and confined 3D configurations. Three types of phenomena, i.e., hydrodynamic, body-force and diffusive-thermal effects that are responsible for the intrinsic instability of premixed flames were investigated in detail, and the interested readers are referred to the review work by Kadowaki and Hasegawa \citep{ kadowaki2005numerical}. Recent works of 2D DNS for thermodiffusively unstable premixed hydrogen flames are briefly reviewed next. Day et al.~\citep{day2011numerical} investigated the interactions of fuel-lean premixed hydrogen/air flame with a weakly turbulent velocity field in a 2D configuration for a range of equivalence ratios focusing on the nitrogen oxide formation. They found that the dominant reaction pathway is changed from NNH to \ce{N2O} as the fuel mixtures become richer (but still at fuel-lean conditions). Altantzis et al.~\citep{altantzis2012hydrodynamic} adopted a single-step chemistry and detailed transport model to study premixed hydrogen/air flames in 2D channel-like domains, aiming at studying the initial linear growth of perturbations as well as the long-term non-linear evolution. They found that the flame dynamics depend strongly on the domain size and on the Lewis number in the non-linear regime. Frouzakis et al.~\citep{frouzakis2015numerical} conducted extensive numerical simulations for hydrogen/air flames to identify the range of dominance regarding the equivalence ratio and domain size for the hydrodynamic instability, the flame shapes, and their propagating speed. Their hydrogen/air mixtures range from rich ($\phi_0 = 2$) to lean ($\phi_0 = 0.5$) conditions, and the height of the 2D domain was varied from 3 to 100 flame thicknesses. They found that the planar laminar flame is unconditionally stable when the domain is narrower than a critical height. For lager domains, the ranges of unstable wavenumbers and the maximum linear growth rate increase with decreasing the equivalence ratio. Yu et al.~\citep{yu2017nonlinear} conducted 2D simulations to investigate cellular instability of lean hydrogen/air laminar premixed flames with an equivalence ratio of 0.6 at elevated pressure conditions ($p_0 = 5 \, \mathrm{atm}$ and 25\,atm). They found two interesting phenomena in the nonlinear evolution process, i.e., mode-lock, and preferential choice of modes. Berger et al.~\citep{berger2019characteristic} conducted a parametric study to investigate the effects of domain size on the flame burning velocity based on a 2D configuration. They found that the total burning velocity of the thermodiffusively unstable flame strongly depends on the domain size and an impact of the domain size on the flame front topology can be obtained in the non-linear phase. With a sufficiently large 2D computational domain (a minimum lateral width of 100 laminar flame thicknesses), a burning velocity independent of the domain size can be obtained, which suggests the existence of a smallest and a largest intrinsic length-scale of the flame front corrugation. In follow-up works, Berger et al.~\citep{berger2022intrinsicPart1, berger2022intrinsic} investigated the impact of intrinsic combustion instabilities on lean hydrogen/air premixed flames at various equivalence ratios ($\phi_0 = 0.4 \sim 1$), unburnt temperatures ($T_0 = 298 \, \mathrm{K} \sim 700 \, \mathrm{K}$), and pressures ($p_0 = 1 \, \mathrm{bar} \sim 20 \, \mathrm{bar}$), focusing on the dispersion relations in the linear regime \citep{berger2022intrinsicPart1} and flame speed enhancement in the non-linear regime \citep{berger2022intrinsic}. They found that the intrinsic instabilities are enhanced with a decrease of equivalence ratio and unburnt temperature and an increase of pressure in both linear and non-linear regimes. Wen et al.~\citep{wen2022flamepart1, wen2022flamepart2} conducted flame structure analyses for spherically expanding thermodiffusively unstable premixed hydrogen flames at atmospheric and elevated pressure conditions. A 2D configuration featuring rotationally symmetrical wedges was used for simulations. They proposed a composition space modeling method to predict the thermodiffusively unstable premixed hydrogen flames, in which the effects of preferential diffusion, strain rate, and curvature are considered in a single set of premixed flamelet equations in composition space \citep{scholtissek2019self1}. It was found that the curvature-sensitive-species and production rates can be accurately predicted by the proposed composition space modeling method for both atmospheric and elevated pressure conditions. Recently, Howarth et al.~\citep{howarth2022empirical} proposed an empirical characteristic scaling model for freely-propagating premixed hydrogen flames by conducting 2D DNS for a wide range of reactant conditions. Based on the DNS dataset, they found that the thermodiffusive response can be well characterized in terms of the second-order instability parameter, which motivates the formulation of the empirical model.

Although 2D DNS can be interesting for parametric studies, as already widely used in the literature, it is clear that 2D DNS cannot accurately calculate the realistic front topology of a realistic case, particularly for strongly curved flames. Thus, 3D DNS is preferred for the accurate calculation of the 3D cellular flame front. Pioneering 3D DNS studies have been conducted for thermodiffusively unstable premixed hydrogen flames \citep{kadowaki1996numerical, kadowaki1999lateral, kadowaki2001body, kadowaki2002formation, day2009turbulence, day2015combined, aspden2011characterization, aspden2011turbulence, aspden2015turbulence, aspden2017numerical, aspden2019towards, altantzis2013numerical, bell2013simulation, wang2018pressure, you2020modelling, lu2021modeling, berger2022synergistic, rieth2023effect, dinesh2016high, song2022diffusive, lee2022dnsPart1, lee2022dns, howarth2022empirical, howarth2023thermodiffusively}, and the recent works are reviewed next. Day et al.~\citep{day2009turbulence} investigated turbulence effects on cellular burning structures in lean premixed hydrogen flames. The geometry of the flame surface was investigated, and the distributions of the mean and Gaussian curvature and the cell sizes were quantified. In a follow-up paper, Day et al.~\citep{day2015combined} proposed a Lagrangian pathline approach to avoid the difficulty in defining a proper flame surface in regions with little or no burning. Aspden et al.~\citep{aspden2011characterization} performed 3D simulations of thermodiffusively unstable premixed turbulent flames over a range of equivalence ratios and the obtained data were used to define modified Karlovitz and Damk{\"o}hler numbers. They found that the new definitions can effectively eliminate the dependence on fuel equivalence ratio for turbulent flames. Aspden and co-workers \citep{aspden2011turbulence, aspden2015turbulence, aspden2017numerical, aspden2019towards} also conducted a series of other relevant works focusing on different aspects in turbulent combustion, e.g., the transition to the distributed burning regime \citep{aspden2011turbulence, aspden2019towards}, turbulence-chemistry interaction \citep{aspden2015turbulence}, diffusive effects \citep{aspden2017numerical}, etc. Altantzis et al.~\citep{altantzis2013numerical} studied the evolution of expanding, lean ($\phi_0 = 0.6 $), premixed hydrogen/air flames in a 3D cylindrical configuration using a single-step reaction and detailed transport. For the 3D cylindrical sector considered, the length in the third (axial) direction is set to be 20 laminar flame thicknesses. They found that the 3D flame experiences negative displacement speeds relative to the flow along positively-curved flame segments due to a wider range of curvatures compared with 2D. Bell et al.~\citep{bell2013simulation} investigated the role of thermodiffusive instabilities on nitrogen emissions in a laboratory-scale low swirl burner fueled with a lean hydrogen-air mixture at atmospheric pressure. The chemical pathways that lead to the NO and NO$_2$ formation were quantified, and the reason for the enhancement of the NO$_\mathrm{x}$ emissions were clarified.  You and Yang \citep{you2020modelling} conducted DNS for turbulent premixed hydrogen/air flames at an equivalence ratio of 0.6, an unburnt temperature of 300\,K, and atmospheric pressure. The domain size in the crosswise direction was around 5.5 times the laminar flame thickness. The DNS dataset was used to evaluate the performance of the proposed model in predicting the turbulent burning velocity. They found that the predictions of the model based on Lagrangian statistics of propagating surfaces agree well with DNS in a wide range premixed combustion regimes. With a similar domain size in the lateral direction, Lu and Yang \citep{lu2021modeling} investigated the pressure effects on the turbulent burning velocity for various pressures and turbulence intensities. It was found that the turbulent burning velocity and flame surface area are influenced by the stretch factor at elevated pressures. They proposed a predictive model for the turbulent burning velocity by combining sub-models of the stretch factor and flame surface area, and good agreements between model predictions and DNS results were obtained. Berger et al.~\citep{berger2022synergistic} conducted large-scale 3D DNS for thermodiffusively unstable premixed hydrogen flames in turbulent and laminar conditions. The turbulent flame is stabilized in a slot burner configuration at the jet Reynolds number of 11000 and Karlovitz number of 15, while the laminar flame is in a cuboid configuration with the largest domain size being 34 laminar flame thicknesses in the crosswise direction. Compared with the thermodiffusively unstable laminar flame at the same conditions, the variations of the local equivalence ratio and local production rates are significantly enhanced in the turbulent flame due to higher fluctuations of curvature and an enhanced average strain rate induced by turbulence. Through detailed analyses, they confirmed that the thermodiffusive instabilities are sustained in turbulent hydrogen flames and show synergistic interactions with turbulence. Lee et al.~\citep{lee2022dnsPart1, lee2022dns} performed DNS to investigate the extreme and leading points in turbulent hydrogen flames focusing on the local thermochemical structure and production rates, and local velocity field and flame topology. Chu et al.~\citep{chu2022effects} conducted 3D DNS for lean hydrogen flame kernels under engine conditions to investigate the effects of preferential diffusion on flame kernel growth and whether thermodiffusive instabilities appear under realistic engine conditions with elevated in-cylinder pressure and high unburned temperature. They found that the strong thermodiffusive instabilities significantly facilitate the flame kernel growth, and also lead to large variations of the local fuel/air equivalence ratio, together with the mechanisms of flame surface area formation. Rieth et al.~\citep{rieth2023effect} conducted 3D DNS to investigate the effect of pressure on the propagation of turbulent fuel-lean premixed hydrogen-air flames. They found that thermodiffusive instability and the ratio of turbulence to laminar burning velocity are amplified for increasing pressure, which was attributed to the combined effects of multi-dimensional geometrical features of the reaction front and flame speed/thickness sensitivity to preferential diffusion. Recently, Howarth et al.~\citep{howarth2023thermodiffusively} conducted various simulations to investigate the influence of reactant conditions on thermodiffusive response in laminar and turbulent flames. The 3D configuration features 28 flame thicknesses in the lateral direction of the domain. They found that the thermodiffusive response can be well characterized by the instability parameter as in 2D, although the model constants are larger. They also found that the turbulence and thermodiffusive instability have an overall similar effects, i.e., increasing curvature, production rates and temperature. Although progress has been made, the characteristic patterns of thermodiffusively unstable premixed hydrogen flames have not yet been studied without confinement effects, and the dependence of the global burning velocity on the 3D domain size have not yet been quantified.

Within the aforementioned context, the motivation of the present work is to perform large-scale DNS for laminar thermodiffusively unstable premixed hydrogen flames in 3D computational domains using a detailed chemical reaction mechanism that includes NO$_\mathrm{x}$ formation. To investigate the effects of confinement on the 3D-flame front evolution, the domain size is reduced from 100 to 17 laminar flame thicknesses in the crosswise and spanwise directions. Around 1.6 billion grid points were used to fully resolve the flame structure in the largest domain. Based on the dataset with the largest 3D domain size, the overall structure of the thermodiffusively unstable premixed flame is first studied, and the effects of confinement and computational setup (2D vs.~3D) on the characteristic patterns are quantified, including the global burning velocity, flame surface area, stretch factor, curvature distribution, and cell size. Then, the distributions of the thermo-chemical quantities in the progress variable and the curvature spaces are analyzed to evaluate the key correlations in the strongly-curved cellular flame structures. Finally, the formation and destruction mechanisms of the characteristic flame front pattern observed in the 3D thermodiffusively unstable flame are discussed. In particular, the contributions of kinematic restoration, curvature dissipation, and production of flame density function to the formation and destruction of flame surface area are quantified. Regarding the NO$_{\mathrm{x}}$ formation chemistry in the 3D thermodiffusively unstable premixed hydrogen flame, a detailed reaction pathway analysis is conducted in Part II of this work \citep{wen2023threepart2}. Furthermore, a new flamelet tabulation model is proposed in Part II to predict the NO$_\mathrm{x}$ formation in thermodiffusively unstable premixed flames, in which a wide range of curvatures associated with the strongly corrugated flame fronts is incorporated in the flamelet table. 

The remainder of the paper is organized as follows. In Section \ref{Sec:2}, the computational setups of the thermodiffusively unstable hydrogen flame are presented. The mathematical aspects are given in Section \ref{Sec:3}. Results are discussed in Section \ref{Sec:4}, and the last section summarizes the findings.

\section{Computational aspects}
\label{Sec:2}

In this work, thermodiffusively unstable premixed hydrogen flames stabilized in a 3D configuration are simulated. The computational domain is a cuboidal field, see Fig.~\ref{fig:3Dstructure}. The domain size is set to be 100 flame thermal flame thicknesses in the lateral $x$- and $z$-directions ($L_x = L_z$), with the length in the streamwise $y$-direction ($L_z$) 150 thermal flame thicknesses. The flame is initialized in the $x$-$z$ plane. The flame thickness is defined by the maximum gradient of temperature for the corresponding unstretched flame. The largest values of $L_x$ and $L_z$ are chosen according to the findings reported by Berger et al.~\citep{berger2019characteristic} for a 2D configuration where it was found that for the conditions studied here, the largest flame-finger structures are on the order of 50 flame thicknesses, and as a consequence, results became independent for lateral domain sizes larger than about 100 flame thicknesses. Since the size of the flame fingers in the present simulations is found to be of similar size as for the 2D simulations, the same lateral width was chosen here as the largest domain. To quantify the effects of confinement on the flame dynamics, the lengths in the lateral directions are varied, while keeping the length in the streamwise $y$-direction constant. Specifically, $L_x$ and $L_z$ are reduced to be 50, 34, and 17 thermal flame thicknesses of the corresponding 1D unstretched premixed flamelet. The unburnt mixture of hydrogen/air with an equivalence ratio $\phi_0$ of 0.4, a temperature $T_0$ of 298\,K, and a pressure $p_0$ of 1\,atm flows into the domain through the $x$-$z$ plane at $y=0$. For the corresponding 1D freely-propagating premixed flame, the laminar flame speed $s_L$ is calculated to be 0.2\,m/s, the flame thickness $l_F$ is 0.62\,mm, and the adiabatic flame temperature $T_{ad}$ is 1421\,K using the FlameMaster package \citep{pitsch1998flamemaster}. The inlet fluid velocity is set to be constant at 1.044\,m/s, which is around 5 times the laminar flame speed so that the flame remains within the domain for a sufficiently long time interval. During the simulation, the flame front is stabilized sufficiently far from the inlet boundary to avoid an impact of inlet on the flame dynamics. The computational domain is initialized with the corresponding 1D freely-propagating premixed flame along the $y$-direction. To trigger the flame instabilities, the flame front is perturbed with a weak harmonic function in the $x$- and $z$-directions, described by $F(y) = A_0 \sin(2 \pi x / \Lambda ) \cdot \sin (2 \pi z / \Lambda) $, see Fig.~S1 in the Supplementary Material. The parameters of $A_0$ and $\Lambda$ are the amplitude and the wavelength of the perturbation, respectively, which are set to be $A_0 = 0.5 l_F$ and $\Lambda = 10 l_F$. Note that the perturbation method, including the parameters of wavelength and the perturbation magnitude, is the same as the 2D study by Berger et al.~\citep{berger2019characteristic} for the same operating conditions. With the same perturbation method, Kadowaki \citep{kadowaki1996numerical} validated that the dispersion relation for the linear regime in the 3D flames is nearly the same as that in the 2D flames, see Figs.~3 and 4 in ref.~\citep{kadowaki1996numerical}. A periodic boundary condition is set in the $x$- and $z$-directions, and an outlet boundary condition is specified at $y = L_y$ for the $x$-$z$ plane. As there are no turbulent velocity fluctuations superimposed on the inlet flow field, and the lateral and outlet boundaries also do not affect the flow field, the increased flow field velocity at around the flame front can only be induced by gas expansion. As the flow field velocity at the flame front becomes constant after the simulation achieves a statistically-steady state, see Fig.~S2 in Section 2 of the Supplementary Material, the flame cannot become turbulent even for a sufficiently long simulation time.

For all four 3D simulations with different domain sizes, the same mesh resolution is applied, and the flame thickness is resolved by at least 10 grid points. Specifically, for the largest domain size, $1024 \times 1536 \times 1024$ grid points are uniformly set in the $x$-, $y$-, and $z$-directions, respectively, resulting in about 1.6 billion grid points. A reduced version of the chemical reaction mechanism developed by Glarborg et al.~\citep{glarborg2018modeling} is adopted, which contains 21 species and 109 elementary reactions. The adopted chemical reaction mechanism is provided in the Supplementary Material. Considering the chemical reaction mechanism, the 3D simulation with the largest domain size results in nearly 34 billion degrees of freedom. The simulations ran on 36864 cores and consumed 67 million CPU-hours for the four simulations to achieve converged results. For the thermodiffusively unstable premixed hydrogen flames, the convergence of the simulation is indicated by the fact that the global burning velocity achieves a statistically-steady state. For the simulation with the largest domain size, the simulation becomes converged after $t = 0.06 \, \mathrm{s}$, i.e., 20 flame times ($t_F = l_F/s_L$). Then, the simulation runs for another 47 flame times to collect statistics for time-averaged analyses.

As a reference case, a 2D simulation is conducted with the domain size of 100 laminar flame thicknesses in the $x$- and $y$-directions using the same mesh resolution and chemical reaction mechanism.

\section{Mathematical aspects}
\label{Sec:3}

\subsection{Governing equations and numerical scheme}
\label{Subsec:31}

In the 3D DNS, the governing equations for the momentum, species mass fractions, and temperature in the low-Mach limit formulation \citep{tomboulides1997numerical} are solved. The diffusion coefficient of each species $\mathscr{D}_k$ is calculated using the thermal conductivity $\lambda$, density $\rho$, specific heat capacity of the mixture $c_p$, and the Lewis number of each species $Le_k$, i.e., $\mathscr{D}_k = \lambda/(\rho c_p Le_k) $. The thermal conductivity of each species is calculated according to Eucken \citep{eucken1913warmeleitvermogen}, while the mixture value is evaluated with the method proposed by Mathur et al.~\citep{mathur1967thermal}. The Lewis number for each species is constant and non-unity, and is calculated from the burnt gas region of the corresponding 1D freely-propagating premixed flamelet. The calculated Lewis numbers are listed in Table \ref{table:Lewis}. The molecular diffusion due to the Soret effect is considered using the reduced thermal diffusion model proposed by Schlup and Blanquart \citep{schlup2019reproducing}. 

\begin{table}[h!]
\caption{Lewis numbers of all species in the chemical reaction mechanism for the 3D simulations conducted.} 
\centerline{\begin{tabular} {c c c c c c c c c c c}
\hline 
	N$_2$ & H$_2$ & H & O & O$_2$ & OH & H$_2$O & HO$_2$ & H$_2$O$_2$ & N & NO \\
	1.1885 & 0.2989 & 0.1785 & 0.6856 & 1.0506 & 0.6981 & 0.7676 & 1.0559 & 1.0627 & 0.7754 & 1.0191 \\
\hline 
	NH & NH$_2$ & HNO & H$_2$NN & NO$_2$ & N$_2$O & NNH & HONO & HONO$_2$ & N$_2$H$_3$ \\
	0.6504 & 0.6635 & 1.0592 & 1.1043 & 1.3845 & 1.3236 & 1.0958 & 1.4338 & 1.6251 & 1.2435 \\
\hline
\end{tabular}}
	\label{table:Lewis}
\end{table}

The semi-implicit finite difference code CIAO \citep{desjardins2008high} is employed to solve the reacting Navier-Stokes equations on the Cartesian grids. The momentum equations are discretized with a fourth-order central difference scheme, while the species and temperature equations are discretized with a fifth-order weighted essentially non-oscillatory scheme (WENO5) \cite{jiang1996efficient}. A second-order Crank-Nicolson scheme \citep{crank1947practical} is utilized for time integration. A time-implicit backward difference method is employed to integrate the chemical source terms, as implemented in the stiff ODE solver CVODE as part of the SUNDIALS suite \citep{hindmarsh2005sundials}. The symmetric operator splitting
method proposed by Strang~\citep{strang1968construction} is utilized to efficiently advance the stiff advection-diffusion-reaction equations for species and temperature. The Poisson equation is solved with the algebraic multi-grid (AMG) solver Hypre Boomer AMG \citep{falgout2002hypre}. To increase numerical accuracy and stability, the spatial and temporal staggering method is utilized in the simulations. 

\subsection{Burning velocity, flame surface area, and stretch factor}
\label{Subsec:32}

Compared with a flat flame, the thermodiffusively unstable premixed flames are  strongly corrugated, which results in increased burning velocity and flame surface area. For a 3D thermodiffusively unstable premixed flame, the global fuel burning velocity $s_c$ can be defined as
\begin{equation}\label{eq:sc}
s_c = - \dfrac{1}{A_0 \rho_u Y_{\mathrm{H_2},u} } \int_{\Xi} \dot{\omega}_{\mathrm{H_2}} \mathrm{d} V,
\end{equation}
where $A_0$ is a reference flame surface area, which is defined as the area of the inlet here, i.e., the area of the $x$-$z$ plane. $\rho_u$ and $Y_{\mathrm{H_2},u}$ are the density and hydrogen mass fraction of the unburnt mixture, respectively. $\dot{\omega}_{\mathrm{H_2}}$ is the consumption rate of the hydrogen molecule, which is integrated over the volume of the whole domain $\Xi$. For the 2D configuration, the global burning velocity is calculated in the same way by replacing the reference flame surface area with the domain size in the crosswise direction $L_x$ and integrating over the area of the entire two-dimensional domain, i.e., $\mathrm{d} V = \mathrm{d}x \mathrm{d} y$. 

The flame surface area $A_\psi$ can be calculated by integrating the instantaneous flame surface density $\sum (\psi, \bm{x}, t ) $ over a volume of $\Xi$ according to Vervisch et al.~\citep{vervisch1995surface}
\begin{equation}\label{eq:area}
A_\psi = \int_{\Xi} \sum (\psi, \bm{x}, t) \mathrm{d} V.
\end{equation}
The flame surface density $\sum (\psi, \bm{x}, t) $ measures the local and instantaneous density per unit volume of the iso-concentration surface area. The flame surface density function of a reactive variable $\varphi (\bm{x},t ) $ at $\varphi (\bm{x},t ) = \psi $ can be obtained as the product of the magnitude of the scalar gradient and the ``fine grained'' PDF \citep{vervisch1995surface}, i.e.,
\begin{equation}\label{surfaceDensity}
\sum \left(\psi, \bm{x},t \right) = |\nabla \varphi \left(  \bm{x},t \right) | \cdot \delta \left[ \psi - \varphi (\bm{x},t ) \right].
\end{equation}
In this work, the iso-surface is defined as $\psi = 1 - Y_{\mathrm{H_2}} / Y_{\mathrm{H_2},u} = 0.8$ to determine the flame surface area following Berger et al.~\citep{berger2022intrinsic} for the same operating conditions. For simplicity, we will denote the flame area $A_{0.8}$ simply as $A$. In ref.~\citep{berger2022intrinsic}, the sensitivity of the choice of species mass fraction for the definition of reactive variable was studied. It was found that the calculated flame surface area is sensitive to the $\psi$-value when it is defined based on the mass fraction of major product $\mathrm{H_2O}$, and an almost constant value can be obtained when $\psi$ is defined based on the $\mathrm{H_2}$ mass fraction. Howarth and Aspden \citep{howarth2022empirical} also reported that the definition of flame surface with the temperature isocontours can significantly overestimate the flame surface area due to the existence of local extinction regions, and found that the definition of flame surface based on the $\mathrm{H_2}$ mass fraction does not experience the same problems. Based on the findings of these works, the reactive variable is defined based on the $\mathrm{H_2}$ mass fraction. To quantify the effects of the choice of isosurface for representation of the flame front in a 3D simulation, the flame surface area is calculated for various values of $\psi$, ranging from $\psi = 0.1$ to $\psi = 0.9$. The results, which are provided in Section 3 of the Supplementary Material, show that the flame surface area calculated with different values of $\psi$ is similar, which confirms that the flame surface area calculation is not sensitive to the choice of the isosurface $\psi$.

The change of burning velocity originates from either the flame wrinkling or the different reactivities along the flame front. The flame wrinkling can be quantified by $A/A_0$, where $A_0$ corresponds to the cross-section of the domain, while the significance of local burning rate can be described by the stretch factor $I_0$, which can be related to the burning velocity and the flame surface area as
\begin{equation}\label{eq:stretchFactor}
I_0 = \dfrac{s_c}{s_L} \left( \dfrac{A}{A_0} \right)^{-1},
\end{equation}
where $s_L$ is the laminar speed of the corresponding 1D freely-propagating premixed flame.

\subsection{Formation and destruction of bulb-like structures}
\label{Subsec:33}

The formation and destruction of the flame surface area can be quantified by the stretch rate $K$, which is defined as
\begin{equation}\label{eq:stretchRate}
K = \dfrac{1}{\delta A} \dfrac{\mathrm{d} \delta A }{\mathrm{d} t},
\end{equation}
where $\delta A$ denotes an element of flame surface area. The stretch rate depends on the curvature of the flame front $\kappa_c$, the flame displacement speed $s_d$, and the tangential strain rate $K_s$ with the relation
\begin{equation}
K  = \kappa_c s_d + K_s.
\end{equation}
As in refs.~\citep{berger2019characteristic, wen2022flamepart1}, the curvature of the flame front $\kappa_c$ is defined as
\begin{equation}
\kappa_c = - \nabla \cdot \bm{n} = - \nabla \cdot \left( \dfrac{ \nabla \varphi }{ | \nabla \varphi | } \right), 
\end{equation}
where $\bm{n}$ is the unit vector in the flame normal direction, and the reactive variable $\varphi = 1 - Y_{\mathrm{H_2}}/Y_{\mathrm{H_2},u}$ is used to define the flame surface, keeping consistency with the calculation of the flame surface area using the formulation defined by Vervisch et al.~\citep{vervisch1995surface}, see Eq.~\eqref{surfaceDensity}. 

The flame displacement speed $s_d$ is determined as \citep{poinsot2005theoretical}
\begin{equation}
s_d = \dfrac{1}{|\nabla \varphi |} \left( \dfrac{\partial \varphi }{\partial t} + \bm{u} \cdot \nabla \varphi \right),
\end{equation}
and the tangential strain rate $K_s$ is calculated as 
\begin{equation}
K_s = \nabla \cdot \bm{u} - \bm{n} \cdot \nabla \left( \bm{u} \cdot \bm{n} \right), 
\end{equation}
where $\bm{u}$ is the gas velocity. 

\section{Results and discussions}
\label{Sec:4}

In Section \ref{Subsec:41}, the characteristic patterns of the cellular flame structure are investigated, and the effects of the computational setup (2D vs.~3D) on the flame dynamics are quantified. In Section \ref{Subsec:43}, the distributions of the thermo-chemical quantities in progress variable space are investigated, focusing on the effects of the computational setup. In Section \ref{Subsec:44}, the correlations between the thermo-chemical quantities and the curvature are quantified by calculating the curvature-conditioned mean values. In Section \ref{Subsec:45}, the formation and destruction mechanism of the distinct bulb-like structure observed in the 3D thermodiffusively unstable premixed flame is discussed, and the contributions to the formation and destruction of the flame surface are quantified.

\subsection{Characteristic patterns}
\label{Subsec:41}

\subsubsection{3D cellular structure}
\label{Subsub:410}

\begin{figure}[!h]
    \centering
    \captionsetup[subfigure]{labelformat=empty}
    \subfloat[]{
    \begin{minipage}[b]{0.45\textwidth}
\tikz[remember picture] \node[inner sep=0pt,outer sep=0pt] (a) 
{\includegraphics[trim = 0mm 0mm 0mm 0mm, clip, angle=0, width=1\linewidth]{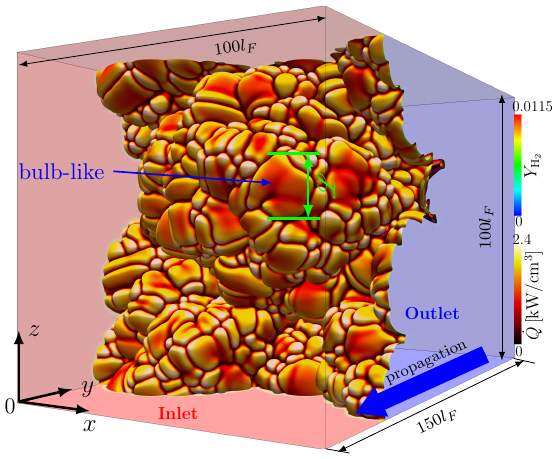}};
\end{minipage}
\begin{tikzpicture}[remember picture,overlay] 
\end{tikzpicture}}
    \vspace{-6mm}
    \caption{\label{fig:3Dstructure} Three-dimensional cellular structure of the thermodiffusively unstable premixed hydrogen flame at $t = 0.2 \, \mathrm{s}$. The iso-surface of $\psi = 0.8 $ is colored by the  local heat release rate. The volume shows the instantaneous distribution of \ce{H2} mass fraction, i.e., the rainbow colorbar. The bulb-like cellular structure and its length-scale $\mathscr{L}$ are indicated. The flame propagating direction and the domain size are also indicated. }
\end{figure}

The overall 3D cellular structure of the thermodiffusively unstable premixed hydrogen flame is visualized in Fig.~\ref{fig:3Dstructure} to give an overall impression. The iso-surface of $\psi = 0.8 $ is colored by the local heat release rate, and the coloring on the domain boundaries shows the instantaneous distribution of the hydrogen mass fraction. In the flame propagating direction, many bulb-like flame structures at various length-scales are formed due to the thermodiffusive instability. Such three-dimensional flame structures obviously cannot be adequately represented in a 2D computational setup. With confined domains, the bulb-like flame structure has also been observed in previous studies \citep{kadowaki2001body, day2009turbulence, bell2013simulation, berger2022synergistic, howarth2023thermodiffusively} for thermodiffusively unstable premixed hydrogen flames operated at similar conditions. However, the statistics of the length-scale of the cell size is impacted by effects of confinement, and will be quantified in Section \ref{Subsubsec:412}. The length-scale of the bulb-like flame front is indicated by $\mathscr{L}$ in Fig.~\ref{fig:3Dstructure}a. The heat release rate is strong for the small cells with sharp edges, while the heat release rate for the large cells has a non-uniform distribution, i.e., the heat release rate is comparably small in the central part of the large cells but large towards the edges.

\subsubsection{2D contour plots}
\label{Subsub:411}

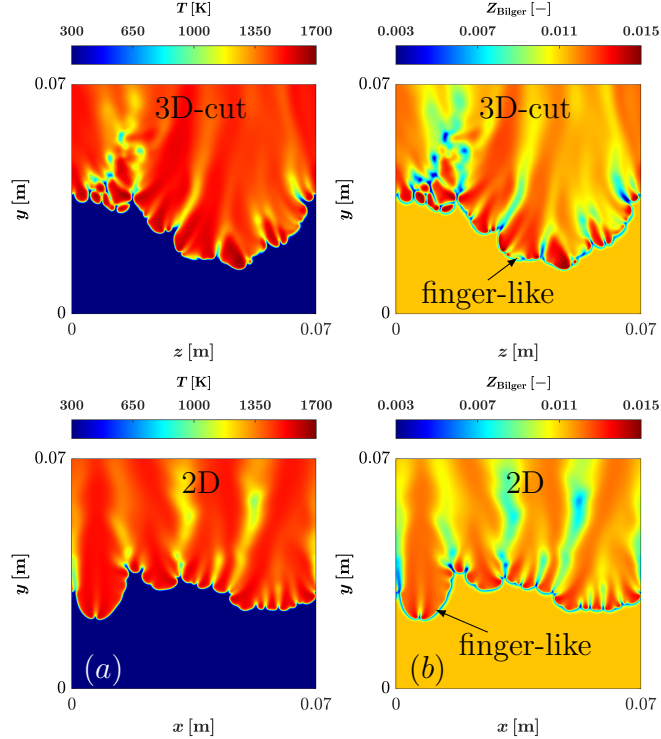
\begin{figure}[!h]
    \centering
    \vspace{-4mm}    
    \captionsetup[subfigure]{labelformat=empty}
    \subfloat[]{\label{subfig:2D3DcutT}
    \begin{minipage}[b]{0.25\textwidth}
\tikz[remember picture] \node[inner sep=0pt,outer sep=0pt] (a) 
{\includegraphics[width=1.02\linewidth]{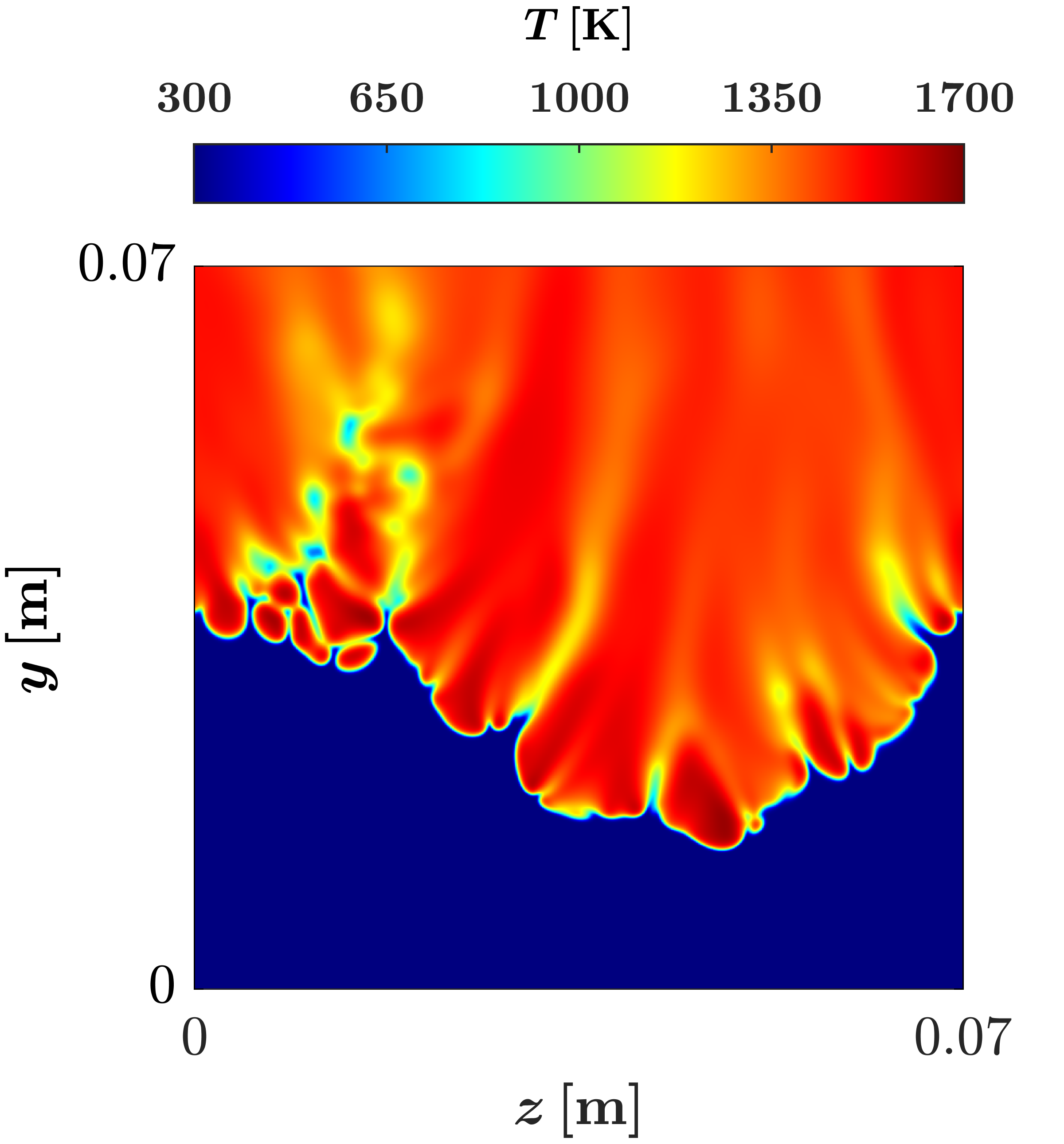}};
\end{minipage}
\begin{tikzpicture}[remember picture,overlay] 
	\node[text=black,scale=1] at (-2,3.4) (a) {$\mathrm{3D}$-cut};	
\end{tikzpicture}} \hspace*{-5mm}
    \subfloat[]{\label{subfig:2D3DcutZBilger}
    \begin{minipage}[b]{0.25\textwidth}
\tikz[remember picture] \node[inner sep=0pt,outer sep=0pt] (a) 
{\includegraphics[width=1.02\linewidth]{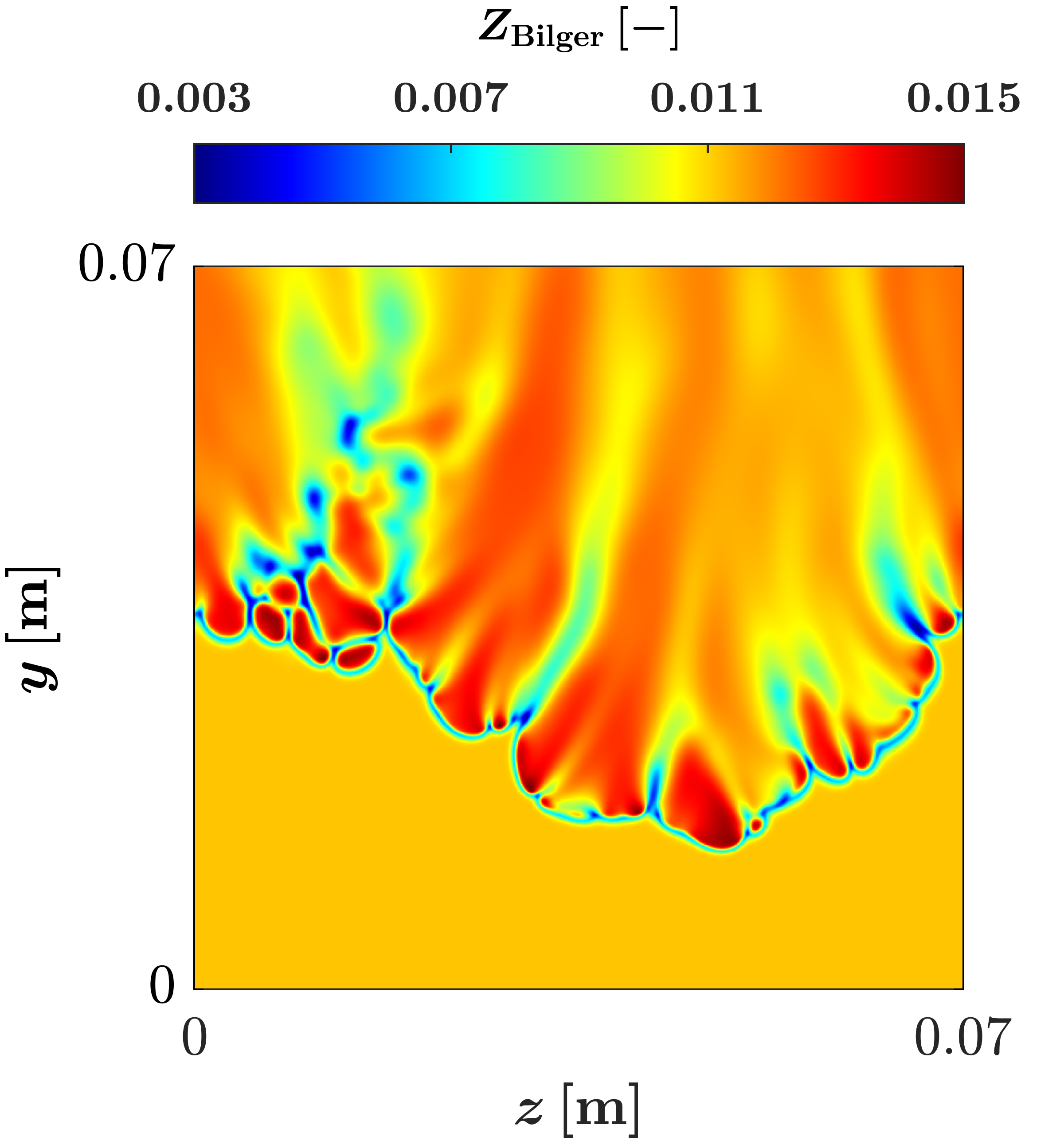}};
\end{minipage}
\begin{tikzpicture}[remember picture,overlay] 
	\node[text=black,scale=1] at (-2,3.4) (a) {$\mathrm{3D}$-cut};	
	\node[text=black,scale=1] at (-2.5,0.9) (a) {finger-like};
    \draw [->,black] (-2.5,1.1) -- (-2.1,1.4);
\end{tikzpicture}}
    \vspace{-10mm}
    \subfloat[]{
    \begin{minipage}[b]{0.25\textwidth}
\tikz[remember picture] \node[inner sep=0pt,outer sep=0pt] (a) 
{\includegraphics[width=1.02\linewidth]{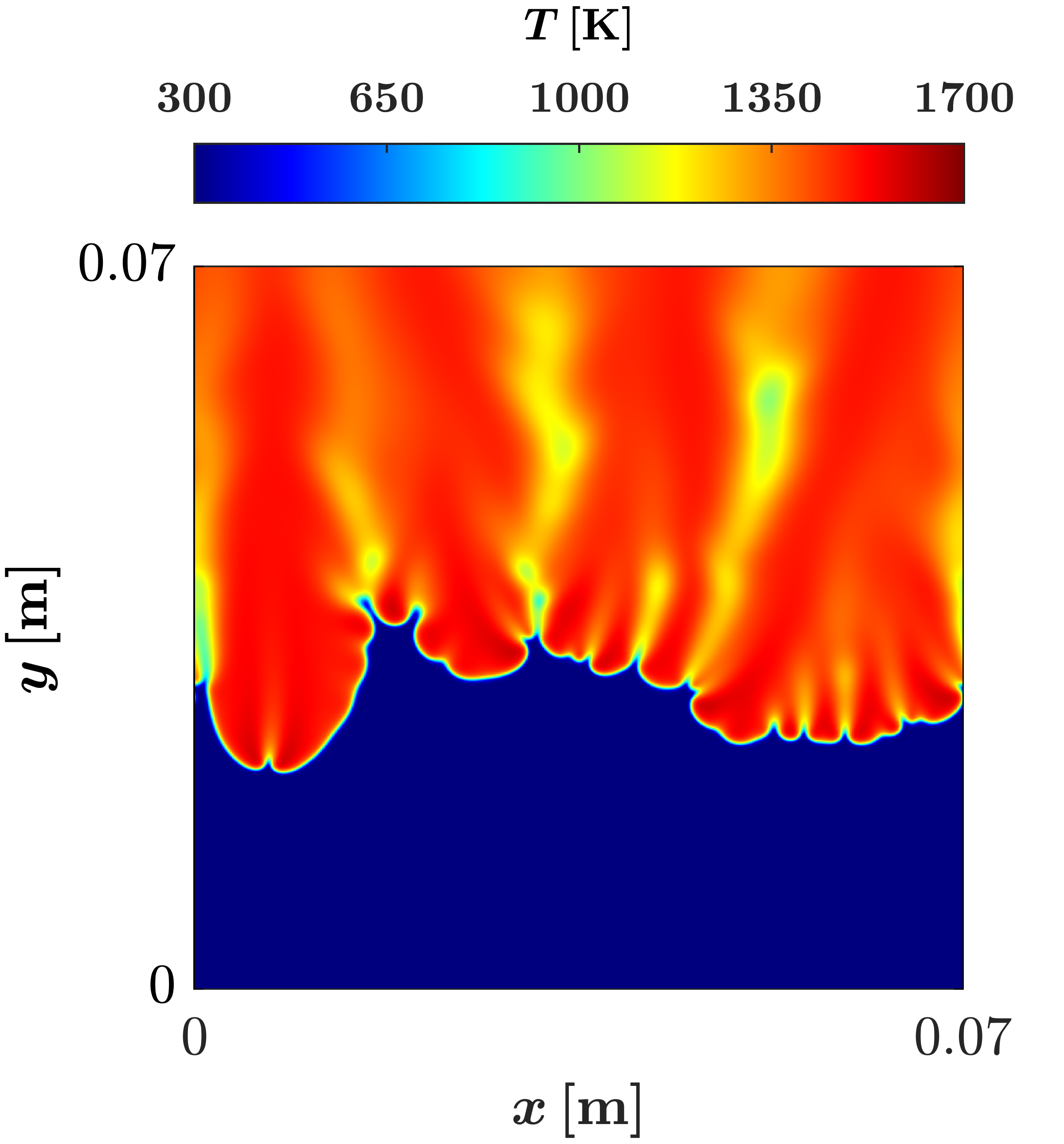}};
\end{minipage}
\begin{tikzpicture}[remember picture,overlay] 
	\node[text=white,scale=1] at (-3.3,0.9) (a) {$(a)$};
	\node[text=black,scale=1] at (-2,3.4) (a) {$\mathrm{2D}$};		
\end{tikzpicture}} \hspace*{-5mm}
    \subfloat[]{
    \begin{minipage}[b]{0.25\textwidth}
\tikz[remember picture] \node[inner sep=0pt,outer sep=0pt] (a) 
{\includegraphics[width=1.02\linewidth]{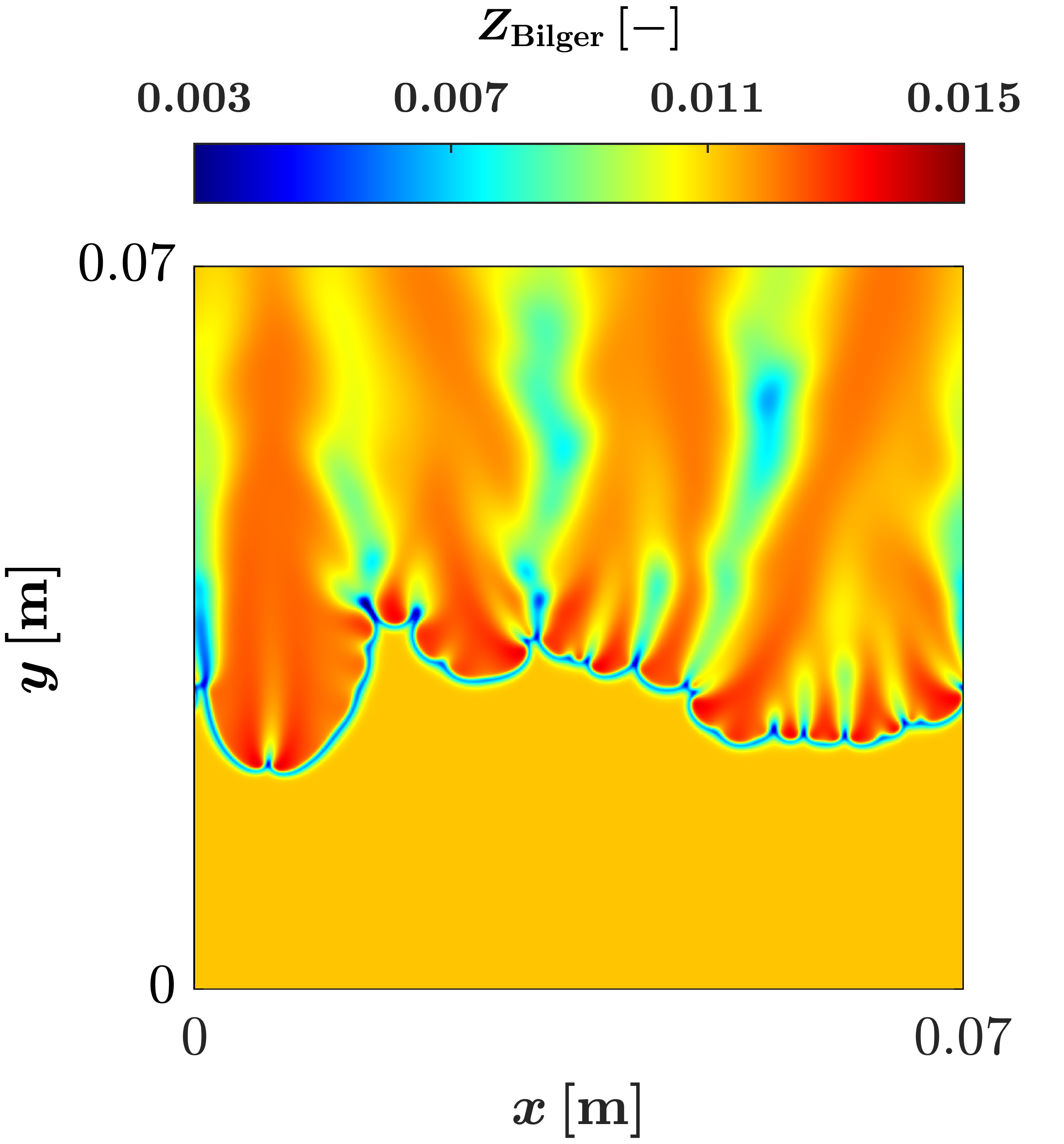}};
\end{minipage}
\begin{tikzpicture}[remember picture,overlay] 
	\node[text=black,scale=1] at (-3.3,0.9) (a) {$(b)$};
	\node[text=black,scale=1] at (-2,3.4) (a) {$\mathrm{2D}$};
	\node[text=black,scale=1] at (-2,1.2) (a) {finger-like};
    \draw [->,black] (-2.5,1.4) -- (-3.2,1.7);	
\end{tikzpicture}}
    \vspace{-8mm}
    \caption{\label{fig:2Dstructure} Instantaneous distributions of the (a) temperature, and (b) Bilger mixture fraction, comparing the $y$-$z$ plane of the 3D simulation (upper row) with the corresponding 2D simulation (lower row). The $y$-$z$ plane of the 3D domain is cut in the $x$ direction to keep the domain size the same as that of the 2D simulation. The finger-like structures in 2D and 3D simulations are indicated.}
\end{figure}

Figure \ref{fig:2Dstructure} shows the instantaneous distributions of the temperature and Bilger mixture fraction for the $y$-$z$ plane of the 3D domain (upper row). For comparison, the results obtained from the corresponding 2D simulation are shown in the lower row. The Bilger mixture fraction is defined based on the elements of H and O as \citep{bilger1990reduced}
\begin{equation}
Z_{\mathrm{Bilger}} = \dfrac{(Z_{\mathrm{H}} - Z_{\mathrm{H},2} )/(2M_{\mathrm{H}}) - (Z_{\mathrm{O}} - Z_{\mathrm{O},2} )/M_{\mathrm{O}} }{ (Z_{\mathrm{H},1} - Z_{\mathrm{H},2} )/(2M_{\mathrm{H}}) - (Z_{\mathrm{O},1} - Z_{\mathrm{O},2} )/M_{\mathrm{O}} },
\end{equation}
where $Z_{\mathrm{H}}$ and $Z_{\mathrm{O}}$ are the local mass fractions of elements H and O, respectively. The subscripts 1 and 2 indicate pure hydrogen and pure air, respectively. Interestingly, although the 3D
cellular structure is completely different from the 2D simulation, the instantaneous distribution
of temperature in 2D and 3D look very similar. For both 2D and 3D simulations, the large-scale finger-like flame structures, as indicated in Fig.~\ref{fig:2Dstructure}, can be observed, both featuring a strong cusp at the flame tip. In addition, many small cells exist along the flame front, indicating the coexistence of different characteristic length-scales of the flame front corrugations in thermodiffusively unstable premixed flames. As shown in Fig.~\ref{subfig:2D3DcutT}, the peak temperatures at the positively-curved flame segments in the 3D simulation are higher than in the 2D simulation, which results from the stronger accumulation of the highly diffusive species in the positively-curved regions. The intensity of the accumulation of the highly diffusive species is associated with the curvature values, which feature larger extreme values in 3D as will be quantified in the next subsection. As indicated by the instantaneous distribution of $Z_{\mathrm{Bilger}}$ in Fig.~\ref{subfig:2D3DcutZBilger}, higher local equivalence ratios are obtained in the positively-curved regions of the 3D simulation compared to the 2D simulation. Similar observations, such as super-adiabatic regions, local enrichment of fuel species, were also reported in previous studies for sub-unity Lewis number laminar flames (e.g., refs.~\citep{day2009turbulence, bell2013simulation, wen2022flamepart1, wen2022flamepart2, howarth2023thermodiffusively}) and turbulent flames (e.g., refs.~\citep{aspden2011characterization, rieth2023effect, rieth2022enhanced, berger2022intrinsic}).

\subsubsection{Effects of confinement}
\label{Subsubsec:412}

\begin{figure}[!h]
    \centering
    \captionsetup[subfigure]{labelformat=empty}
    \subfloat[]{
    \begin{minipage}[b]{0.55\textwidth}
\tikz[remember picture] \node[inner sep=0pt,outer sep=0pt] (a) 
{\includegraphics[width=1.02\linewidth]{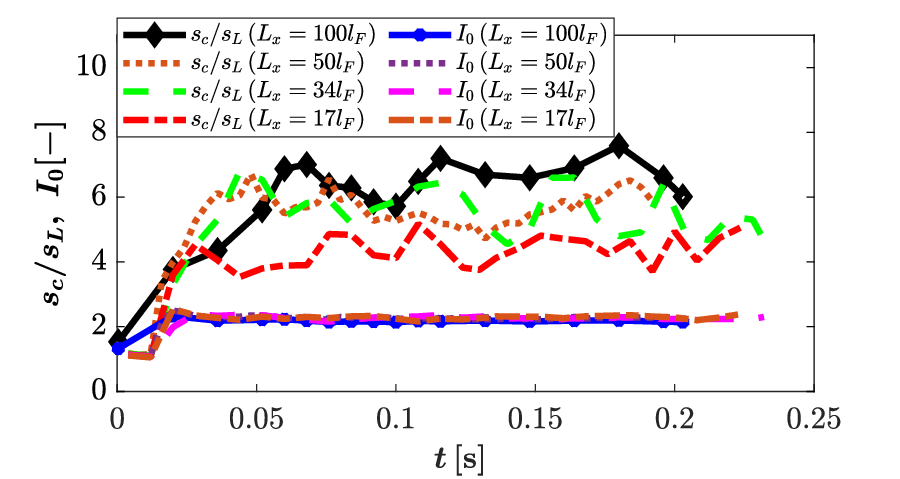}};
\end{minipage}
\begin{tikzpicture}[remember picture,overlay] 
\end{tikzpicture}}
    \vspace{-6mm}
    \caption{\label{fig:scI0DomainSize} Comparison of the normalized global burning velocity and stretch factor for the 3D thermodiffusively unstable premixed hydrogen flame stabilized in different domain sizes.}
\end{figure}

\begin{table}[h!]
\caption{The mean values of the normalized burning velocity, the flame surface area, and the stretch factor obtained from different domain sizes.} 
\centerline{\begin{tabular} {L{6cm} L{2cm} L{2cm} L{2cm} L{2cm}}
\hline 
	 Parameters & $17 l_F$ & $34 l_F$ & $50 l_F$ & $100 l_F$ \\
\hline 
	 Burning velocity $s_c/s_L$ & 4.38 & 5.71 & 5.87 & 6.69 \\
	 Flame area $A/A_0$, $A/L_x$ & 1.87 & 2.44 & 2.51 & 2.86 \\	
	 Stretch factor $I_0$ & 2.34 & 2.34 & 2.34 & 2.34 \\
\hline
\end{tabular}}
	\label{table:domainSize}
\end{table}

As reported by Berger et al.~\citep{berger2019characteristic} for a 2D thermodiffusively unstable premixed  hydrogen flame, the flame dynamics are influenced by the domain size, and a domain-independent burning velocity can be obtained for a sufficiently large domain. To investigate the effects of confinement on the global burning velocity in the 3D simulation, the domain size was varied in the $x$- and $z$-directions from $L_x = L_z = 100 l_F$ to $L_x = L_z = 17l_F$, with a constant length in $y$-direction of $L_y = 150 l_F$. Figure \ref{fig:scI0DomainSize} compares the normalized flame burning velocity $s_c/s_L$ and the stretch factor $I_0$ as defined in Eqs.~\eqref{eq:sc} and \eqref{eq:stretchFactor}, for the 3D thermodiffusively unstable premixed flame calculated with different domain sizes. The mean values are presented in Table \ref{table:domainSize}. It can be observed that a domain-independent behavior cannot be observed when the domain size is not larger than $50 l_F$. Previous simulations from different groups showed that a constant value can be obtained when the domain size in the lateral direction is larger than 100 flame thicknesses, see Fig.~18 in the work of Creta et al.~\citep{creta2020propagation}, which indicates the existence of a largest corrugation scale. For the stretch factor, no effects of the confinement are observed as it stays around 2 indicating that $I_0$ is determined on the small scales. The visualization of the 3D cellular flame structure obtained from different domain sizes is provided in Section 4 of the Supplementary Material. Considering the effects of confinement, the discussions on the 3D DNS in the following sections of this work are only based on the largest domain size with statistically stationary burning velocity.

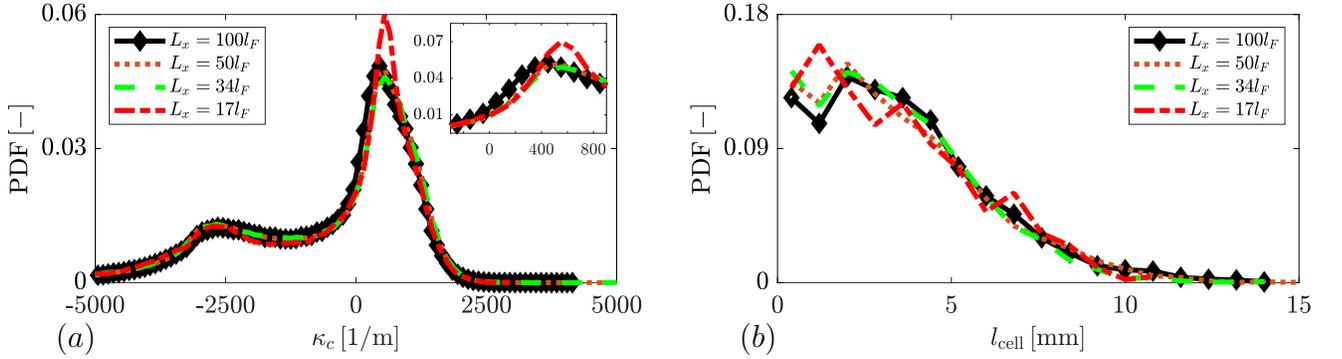
\begin{figure}[!h]
    \centering
    \captionsetup[subfigure]{labelformat=empty}
    \subfloat[]{\label{subfig:kappacSizeA}
    \begin{minipage}[b]{0.5\textwidth}
\tikz[remember picture] \node[inner sep=0pt,outer sep=0pt] (a) 
{\includegraphics[width=1.02\linewidth]{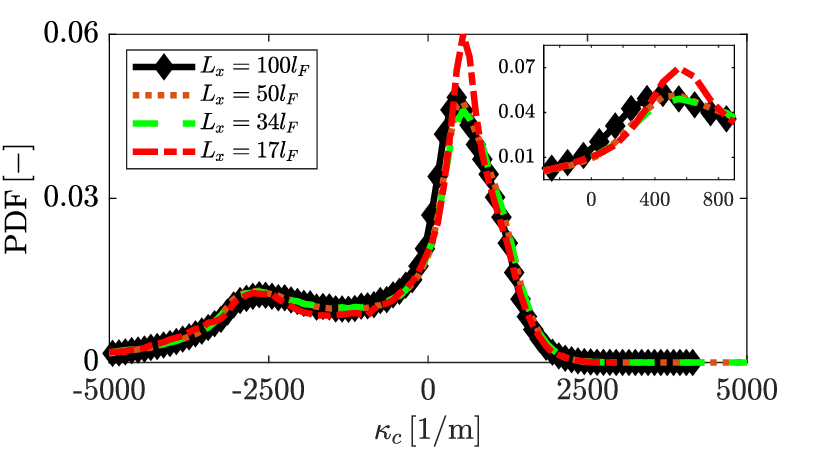}};
\end{minipage}
\begin{tikzpicture}[remember picture,overlay] 
	\node[text=black,scale=1] at (-8,0.2) (a) {$(a)$};
\end{tikzpicture}}
    \subfloat[]{\label{subfig:kappacSizeB}
    \begin{minipage}[b]{0.5\textwidth}
\tikz[remember picture] \node[inner sep=0pt,outer sep=0pt] (a) 
{\includegraphics[width=1.02\linewidth]{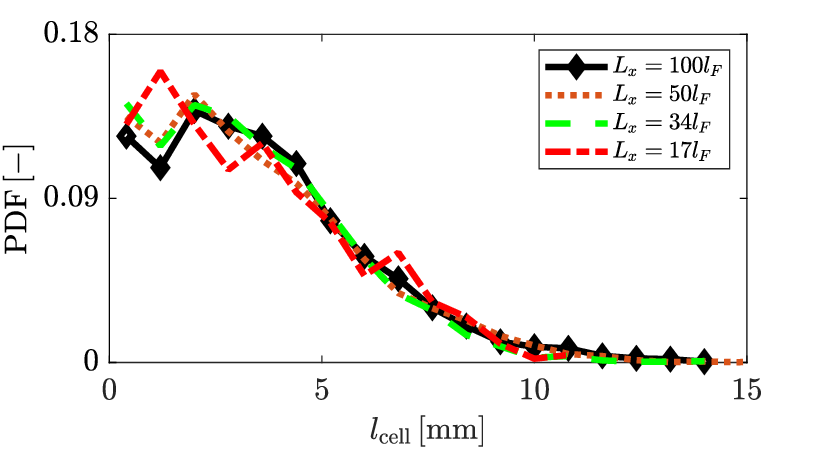}};
\end{minipage}
\begin{tikzpicture}[remember picture,overlay] 
	\node[text=black,scale=1] at (-8,0.2) (a) {$(b)$};
\end{tikzpicture}}
    \vspace{-8mm}
    \caption{\label{fig:kappacCellsizeDomainSize} Comparison of PDF of (a) curvature, and (b) cell size for the 3D thermodiffusively unstable premixed hydrogen flame stabilized in different domain sizes. The curvature statistics are extracted from the whole domain for various time instants, while the cell size is calculated at the slices. The inset in (a) shows a zoom of curvature PDF profiles around the peak. }
\end{figure}

To investigate the effects of confinement further, the PDFs of the curvature calculated from different domains are compared in Fig.~\ref{subfig:kappacSizeA}. The statistics for the smaller domain sizes are extracted from various time instants, i.e., $t = 0.1 \sim 0.22 \, \mathrm{s}$ for the $L_x = 17 l_F$ case, $t = 0.1 \sim 0.232 \, \mathrm{s} $ for the $L_x = 34 l_F$ case, and $t = 0.1 \sim 0.192 \, \mathrm{s} $ for the $L_x = 50 l_F$ case. Small curvatures in the range of $0 < \kappa_c < 400 \, \mathrm{m^{-1}}$ are more prominent in the largest domain case, while large curvatures in the range of $400 \, \mathrm{m^{-1}} < \kappa_c < 800 \, \mathrm{m^{-1}}$ have higher probability in the smallest domain case. Hence, the larger cell size is more prominent in the largest domain, while the smaller cell size dominates in the smallest domain, as shown in Fig.~\ref{subfig:kappacSizeB} of the cell size PDF. The cell size $l_{\mathrm{cell}}$ is defined as the arc length $l_{\mathrm{arc}}$ between two neighboring cusps with negative curvatures for the slices from the 3D domain. The arc length is defined as $l_{arc}^{(n+1)} = l_{arc}^{(n)} + \sqrt{(\mathrm{d} x)^2 + (\mathrm{d} y)^2 } $, where $n$ is the point index along the iso-line of $\psi = 0.8 $. The fractal dimension analysis conducted by Chatakonda et al.~\citep{chatakonda2013fractal} reveals that the global burning rate is related to the ratio of the smallest and the largest characteristic length scales of the flame front corrugation, which explains the increased burning velocity in the largest domain. Thus, the saturation of the flame surface area and burning velocity is associated with the fact that the smallest and the largest characteristic length scales do not change anymore as the domain size becomes sufficiently large.

\subsubsection{Effects of computational setup}
\label{Subsubsec:413}

\begin{figure}[!h]
    \vspace{-3mm}
    \centering \hspace*{-5mm}
    \captionsetup[subfigure]{labelformat=empty}
    \subfloat[]{\label{subfig:scI02D3DA}
    \begin{minipage}[b]{0.5\textwidth}
\tikz[remember picture] \node[inner sep=0pt,outer sep=0pt] (a) 
{\includegraphics[width=1.05\linewidth]{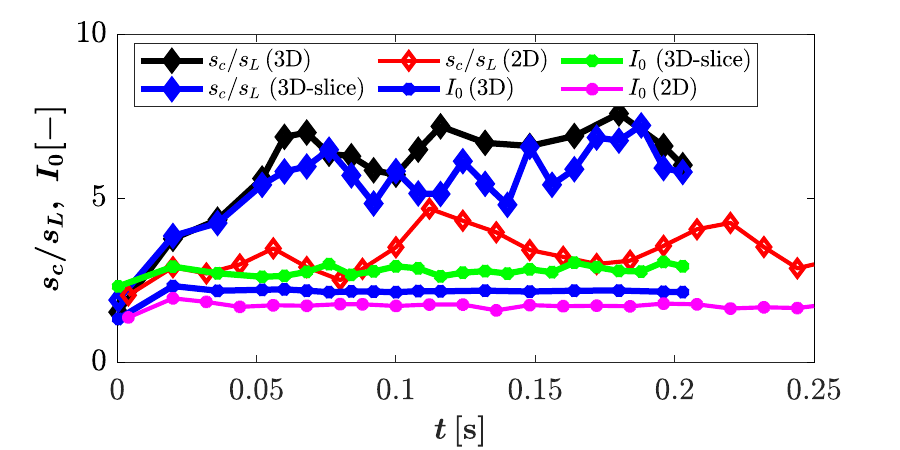}};
\end{minipage}
\begin{tikzpicture}[remember picture,overlay] 
	\node[text=black,scale=1] at (-8,0.2) (a) {$(a)$};
\end{tikzpicture}}
    \subfloat[]{\label{subfig:scI02D3DB}
    \begin{minipage}[b]{0.5\textwidth}
\tikz[remember picture] \node[inner sep=0pt,outer sep=0pt] (a) 
{\includegraphics[width=1.05\linewidth]{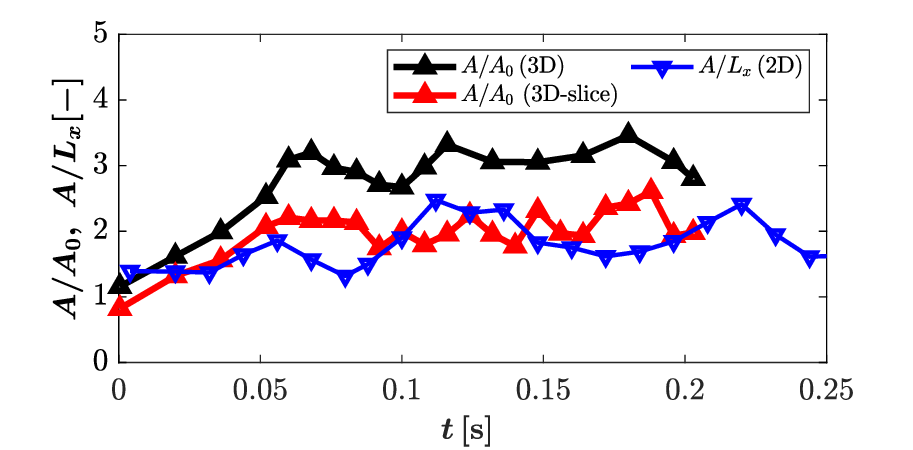}};
\end{minipage}
\begin{tikzpicture}[remember picture,overlay] 
	\node[text=black,scale=1] at (-8,0.2) (a) {$(b)$};
\end{tikzpicture}}
    \vspace{-8mm}
    \caption{\label{fig:scI02D3D} Comparison of (a) flame burning velocity and stretch factor, and (b) flame surface area among the 2D, 3D simulations, and the slices of the 3D simulations for the thermodiffusively unstable premixed hydrogen flames. }
\end{figure}
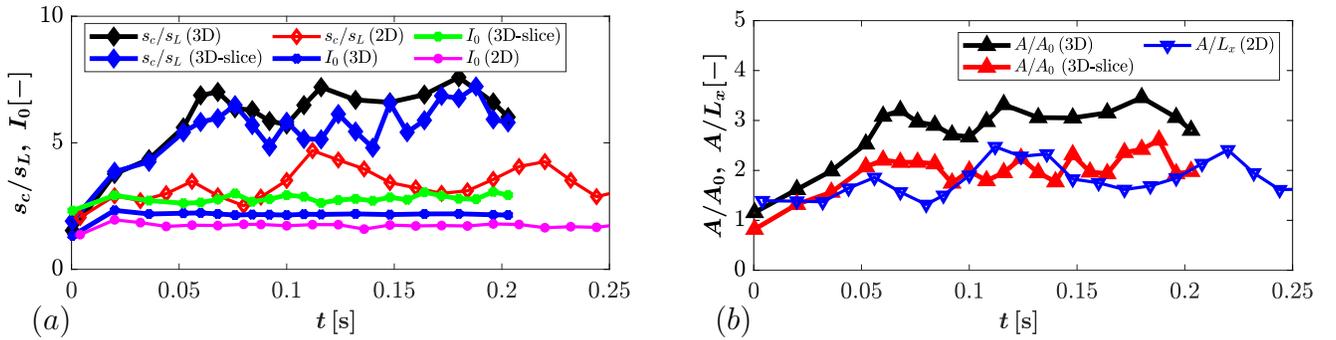

The effects of computational setup (2D vs.~3D) on the flame burning velocity and stretch factor are shown in Fig.~\ref{subfig:scI02D3DA}. The burning velocity of the 3D premixed flame is twice that of the corresponding 2D flame, while the stretch factor is increased by only $28 \%$ in the 3D configuration. Thus, the increased burning velocity in the 3D configuration is predominantly linked to the increased flame surface area\footnote{Note that in the 2D DNS, $s_c/s_L = 3.42$, which is different from the value of 4 in the work by Berger et al.~\citep{berger2019characteristic}. This is associated with the different laminar burning velocities calculated with different chemical reaction mechanisms. However, we note that the differences of laminar flame speed calculated with the different chemical reaction mechanisms are overall small, as presented in Section 5 of the Supplementary Material.}. Similar findings were also reported by Kadowaki \citep{kadowaki2002formation} for much smaller domain with values of 2 for the burning velocity increase and 1.1 for $I_0$. Howarth et al.~\citep{howarth2023thermodiffusively} reported for the same conditions an $I_0$ value of 1.66, but did not evaluate the surface area increase. The comparisons of the flame surface areas of the 2D and 3D configurations are shown in Fig.~\ref{subfig:scI02D3DB}. Note that the flame surface area $A$ for the 2D configuration has a unit of length, which is normalized by the width of the domain in the crosswise direction $L_x$, which is different from the calculation for the 3D configuration. With the difference in flame surface area calculation in mind, the normalized flame surface area in the 3D configuration is around 1.5 times the value in the 2D configuration. 

It is interesting to investigate whether the burning velocity and the stretch factor in the slices of the 3D configuration can be approximated by the corresponding 2D simulation. To this end, Fig.~\ref{subfig:scI02D3DA} also compares the flame burning velocity and stretch factor obtained from the 2D simulation with the slices of the 3D simulation. For both burning velocity and stretch factor, the 2D simulation overall gives under-predictions as time evolves. In addition, the ratio of the normalized burning velocity between the 3D and 2D simulations is close to the ratio of the stretch factor. Figure \ref{subfig:scI02D3DB} compares the ratio of flame surface area between the 3D-slices and the 2D configuration for various time instants. It can be observed that for all time instants, the value of $A/L_x$ calculated from the 3D-slices and the 2D configuration is close. The surface area increase can be explained purely geometrically as a 3D bulb has two positive principal curvatures, whereas for the same structure in 2D there is only one. 

\begin{table}[h!]
\caption{The mean values of the normalized burning velocity, the flame surface area, and the stretch factor obtained from 2D, 3D domains, and the slices of the 3D simulation.} 
\centerline{\begin{tabular} {L{6cm} L{3cm} L{3cm}  L{3cm}}
\hline 
	 Parameters & 2D domain & 3D domain & 3D-slices \\
\hline 
	 Burning velocity $s_c/s_L$ & 3.42 & 6.69 & 5.88 \\
	 Flame area $A/A_0$, $A/L_x$ & 1.90 & 2.86 & 2.08 \\	
	 Stretch factor $I_0$ & 1.80 & 2.34 & 2.83 \\
\hline
\end{tabular}}
	\label{table:asymptotic}
\end{table}

The mean values of the normalized burning velocity, the flame surface area, and the stretch factor for the 2D, 3D domains and the slices of the 3D domain are summarized in Table \ref{table:asymptotic}, which are obtained by averaging the values for the steady condition.

\begin{figure}[!h]
    \centering
    \captionsetup[subfigure]{labelformat=empty}
    \subfloat[]{
    \begin{minipage}[b]{0.5\textwidth}
\tikz[remember picture] \node[inner sep=0pt,outer sep=0pt] (a) 
{\includegraphics[width=1.05\linewidth]{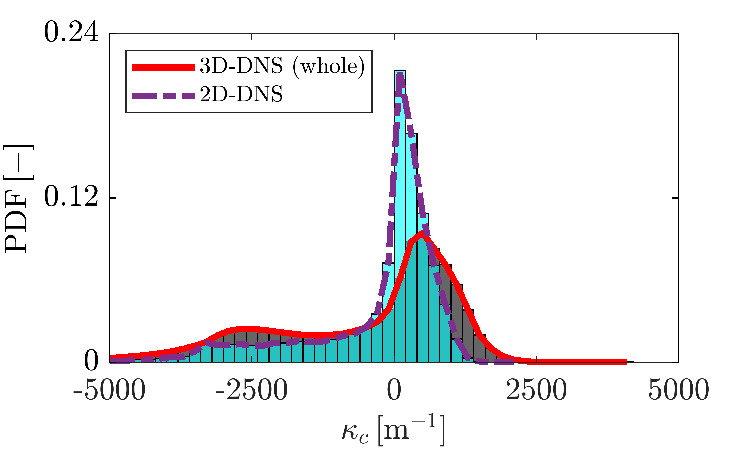}};
\end{minipage}
\begin{tikzpicture}[remember picture,overlay] 
\end{tikzpicture}}
    \vspace{-8mm}
    \caption{\label{fig:curvaturePDF} Comparison of PDF of curvature between the 2D simulation and the 3D simulation. The statistics for the 2D simulation are obtained from sixty time instants. The data for the 3D simulation are extracted from the whole domain.}
\end{figure}

The increased reactivity in the 3D simulation is considered to be directly related to the increased local equivalence ratio and peak temperature, which suggests a stronger accumulation of the fuel. As the curvature of the flame front promotes the effects of preferential diffusion and hence the accumulation of hydrogen, it is essential to investigate the PDF of the curvature in the 2D and 3D simulations. The PDF results obtained from the 2D and 3D simulations are shown in Fig.~\ref{fig:curvaturePDF}. For the 2D simulation, the statistics are collected from $t = 0.252 \, \mathrm{s}$ to $t = 0.488 \, \mathrm{s}$ (78 flame times) equidistantly with a time separation of $0.004 \, \mathrm{s}$ (1.3 flame times). For the 3D simulation, the statistics correspond to the last field. The statistics at the other time instants do not show obvious differences, see Section 6 of the Supplementary Material. For both 2D and 3D simulations, the peak probability appears for the positive curvatures, but the curvature value of the 3D simulation at the peak probability is larger than for the 2D simulation. In addition, the probability of large positive curvature values is higher in the 3D simulation compared to the 2D simulation. This can also be explained purely geometrically as a 3D bulb has two positive principal curvatures, whereas for the same structure in 2D there is only one. The larger 3D positive curvature values induce an accumulation of highly diffusive species, which promotes the reactivity. For the negative curvature, the PDF distribution is overall similar between the 2D and 3D simulations, with only a slight difference at around $\kappa_c = -2500 \, \mathrm{m^{-1}}$. The comparisons of the PDFs of curvature and cell size between the 2D slices of the 3D domain and the 2D simulation are provided in Section 7 of the Supplementary Material.

\subsection{Distributions of thermo-chemical variables in progress variable space}
\label{Subsec:43}

In this subsection, the distributions of the thermo-chemical variables are presented in the progress variable space. In particular, the effects of computational setup and thermodiffusive instability on the distributions of the thermo-chemical variables are quantified. The distributions of the thermo-chemical quantities along the flame normal direction are provided in Section 8 of the Supplementary Material.

\begin{figure}[!h]
    \centering
    \captionsetup[subfigure]{labelformat=empty} \hspace*{-10mm}
    \subfloat[]{
    \begin{minipage}[b]{1\textwidth}
\tikz[remember picture] \node[inner sep=0pt,outer sep=0pt] (a) 
{\includegraphics[width=1.05\linewidth]{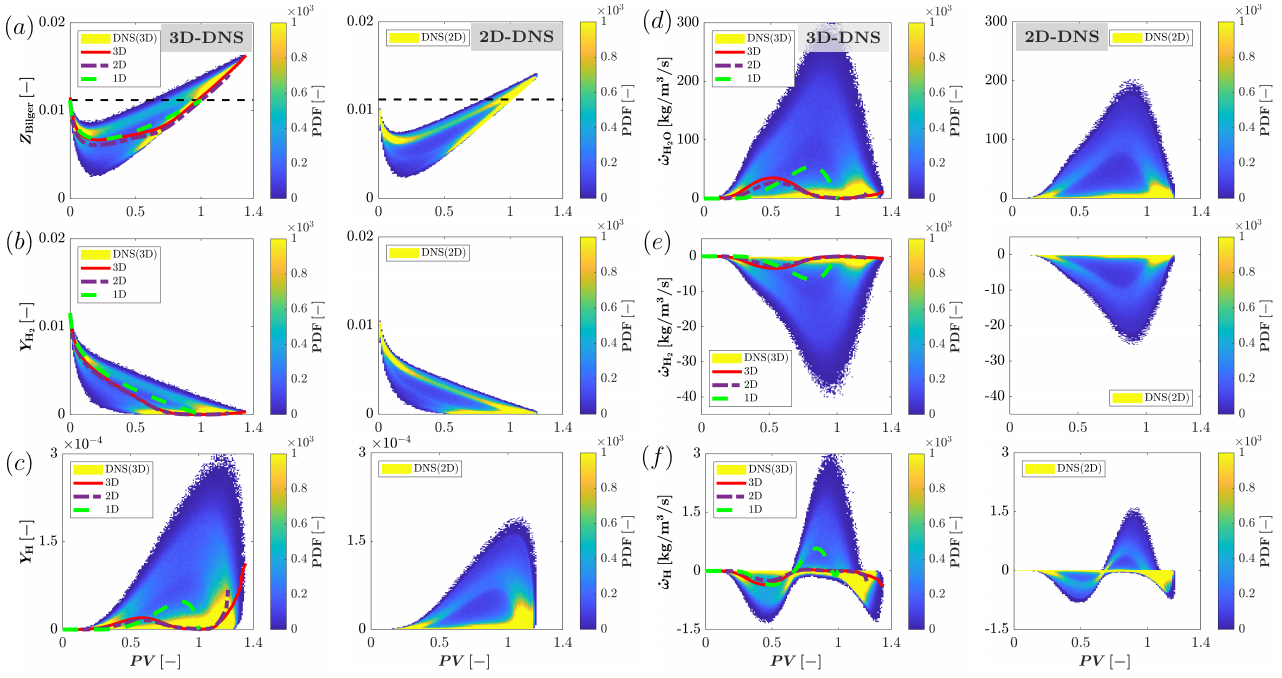}};
\end{minipage}
\begin{tikzpicture}[remember picture,overlay] 
\end{tikzpicture}}
    \vspace{-8mm}
    \caption{\label{fig:condiZBilgerH2H} Comparison of the conditioned mean (a) Bilger mixture fraction, (b) \ce{H2} mass fraction, (c) H radical mass fraction, (d) \ce{H2O} production rate, (e) \ce{H2} production rate, and (f) H radical production rate in progress variable space between the 3D (left column) and 2D (right column) simulations. The results obtained from the corresponding 1D freely-propagating premixed flame are superimposed for comparison. The sample points for the 3D simulation are from the whole domain, while those for the 2D simulation are extracted from sixty time instants to ensure sufficient statistics. The horizontal line in (a) corresponds to the Bilger mixture fraction of the unburnt hydrogen/air mixture. }
\end{figure}

Figures \ref{fig:condiZBilgerH2H}a-\ref{fig:condiZBilgerH2H}c compare the conditioned values of Bilger mixture fraction, \ce{H2} mass fraction, and H radical mass fraction in the normalized progress variable space between the 3D simulation (left column) and the 2D simulation (right column). The normalized progress variable, $PV$, is defined as
\begin{equation}\label{eq:PV}
PV = \dfrac{Y_{PV} - Y_{PV,u}}{Y_{PV,b} - Y_{PV,u}},
\end{equation}
where the subscripts $u$ and $b$ represent the unburnt and burnt states of the corresponding 1D freely-propagating premixed hydrogen flame, respectively. Following previous works for premixed hydrogen combustion \citep{scholtissek2019self1, wen2022flamepart1, wen2022flamepart2, bottler2020premixed}, $Y_{PV}$ is defined based on the major reactants and products, i.e., $Y_{PV} = Y_{\mathrm{H_2O}} - Y_{\mathrm{H_2}} - Y_{\mathrm{O_2}}$. The results obtained from the corresponding 1D freely-propagating premixed flame are superimposed for comparison. The symbol $\left\langle \cdot \right\rangle$ denotes the spatially-averaged values. The horizontal line in Fig.~\ref{fig:condiZBilgerH2H}a indicates the value of $Z_{\mathrm{Bilger}}$ for the unburnt hydrogen/air mixture. It can be observed that the thermodiffusive instability promotes the accumulation of fuel mixture due to preferential diffusion. Furthermore, the peak value of $Z_{\mathrm{Bilger}}$ in the 3D simulation is larger compared to the 2D simulation, which is consistent with the findings reported in the previous section where higher reactivity was observed for the 3D configuration compared with the 2D domain. It is quite interesting to observe that the conditional averages for mixture fraction, fuel mass fraction, and hydrogen radical mass fraction of the 2D and 3D simulations are very similar. Differences are only seen in the range, where the 3D simulations extend to substantially higher values in mixture fraction and progress variable. This extended range causes the higher reactivity in the 3D simulations. The same behavior is also seen in the chemical source terms for $\mathrm{H_2O}$, $\mathrm{H_2}$, and H, as shown in Figs.~\ref{fig:condiZBilgerH2H}d-\ref{fig:condiZBilgerH2H}f. While differences can be observed in the joint PDFs, the conditional mean values are very close with the 3D results showing an extended range with high probability. This implies that the manifold observed in the 3D simulations can be well described from 2D simulations as long as high enough curvature values are present to form mixtures as rich as those observed in the 3D simulations. This could, for instance, be accomplished by adding shear to the 2D simulations.

\subsection{Distributions of thermo-chemical variables in curvature space}
\label{Subsec:44}

\begin{figure}[!h]
    \centering   
    \captionsetup[subfigure]{labelformat=empty} \hspace*{-10mm}
    \subfloat[]{
    \begin{minipage}[b]{1\textwidth}
\tikz[remember picture] \node[inner sep=0pt,outer sep=0pt] (a) 
{\includegraphics[width=1.05\linewidth]{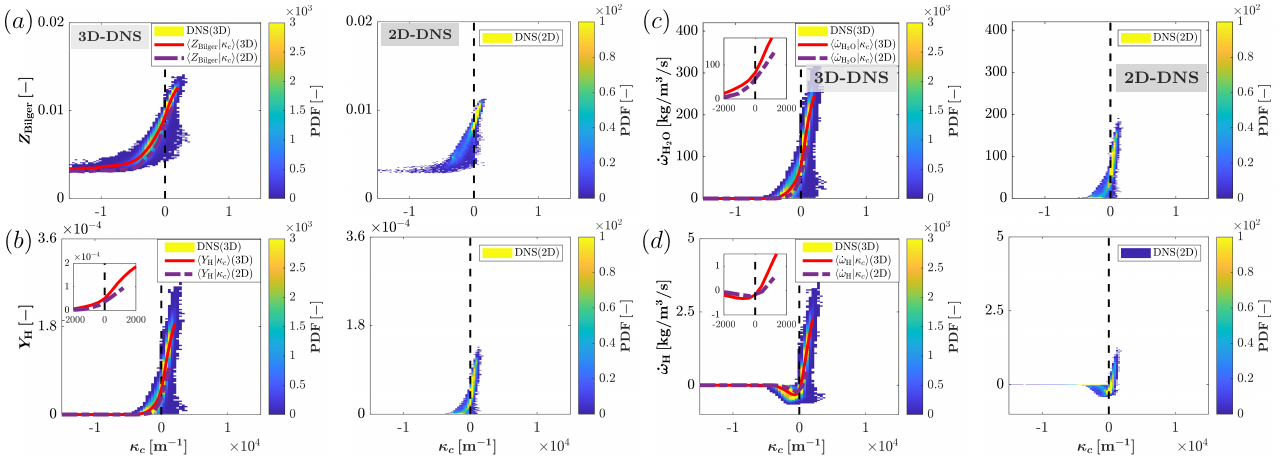}};
\end{minipage}
\begin{tikzpicture}[remember picture,overlay] 
\end{tikzpicture}}
    \vspace{-6mm}
    \caption{\label{fig:ZBilgerTHkappa} Comparison of the conditioned (a) Bilger mixture fraction, (b) H radical mass fraction, (c) \ce{H2O} production rate, and (d) H radical production rate in curvature space between the 3D (left column) and 2D (right column) simulations. The sample points for the 3D simulation are from the whole domain for a given time instant, while those for the 2D simulation are extracted from sixty time instants to ensure sufficient statistics. All the sample points are conditioned on the region of $0.79 < \psi < 0.81 $ around the flame front. The inset shows the profiles of the conditioned mean values at around the zero curvature. }
\end{figure}

For the 3D thermodiffusively unstable premixed hydrogen flame, the distributions of the thermo-chemical quantities are correlated with the length-scale of the cells, which is closely related to the local curvature values $\kappa_c$. $\kappa_c$ is significant at the sharp edges of the cells and is particularly pronounced in regions where negative-curvature lines meet to represent the corners of the bulb-like structures. 

Figure \ref{fig:ZBilgerTHkappa} shows the curvature-conditioned mean values of the Bilger mixture fraction, H radical mass fraction, $\mathrm{H_2O}$ production rate, and H radical production rate for the 3D (left column) and 2D (right column) simulations. The range of curvature in the 3D simulation is wider than the 2D simulation, which is consistent with the observation in Fig.~\ref{fig:curvaturePDF}. The conclusions from the previous section show that the flame structure between 2D and 3D is the same with 3D flames showing a larger mixture fraction range. This implies that the curvature distribution shows similar features, i.e., that the joint scalar and curvature PDFs are similar between 2D and 3D, but larger curvatures exist in 3D that correlate with larger mixture fraction values.

\subsection{Formation and destruction of bulb-like structures}
\label{Subsec:45}

\begin{figure}[!h]
    \centering
    \vspace{-3mm}    
    \captionsetup[subfigure]{labelformat=empty}
    \subfloat[]{\label{subfig:stretch}
    \begin{minipage}[b]{0.33\textwidth}
\tikz[remember picture] \node[inner sep=0pt,outer sep=0pt] (a) 
{\includegraphics[width=1\linewidth]{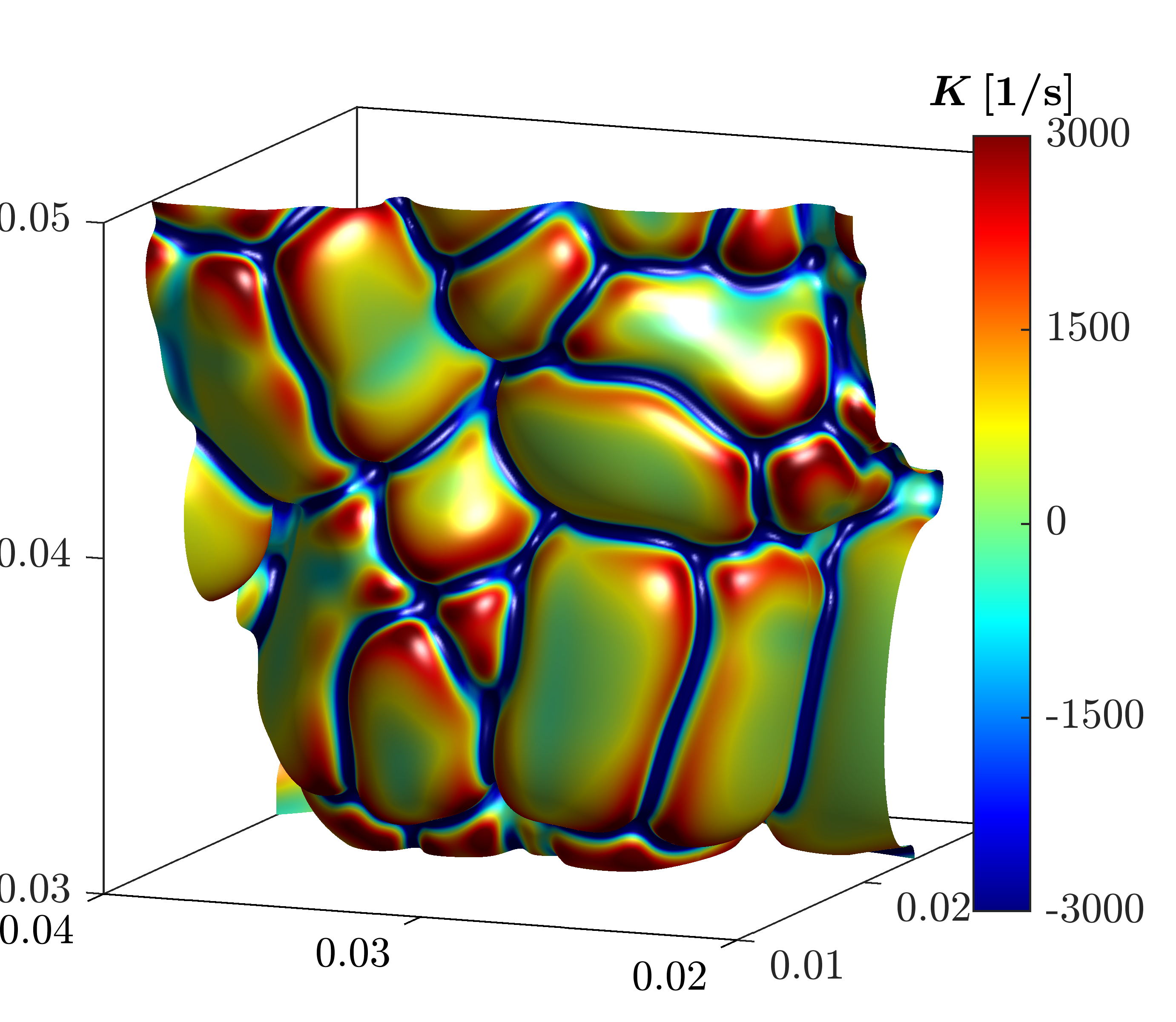}};
\end{minipage}
\begin{tikzpicture}[remember picture,overlay] 
	\node[text=black,scale=1] at (-5.6,0.1) (a) {$(a)$};
\end{tikzpicture}}
    \subfloat[]{\label{subfig:strainRate}
    \begin{minipage}[b]{0.33\textwidth}
\tikz[remember picture] \node[inner sep=0pt,outer sep=0pt] (a) 
{\includegraphics[width=1\linewidth]{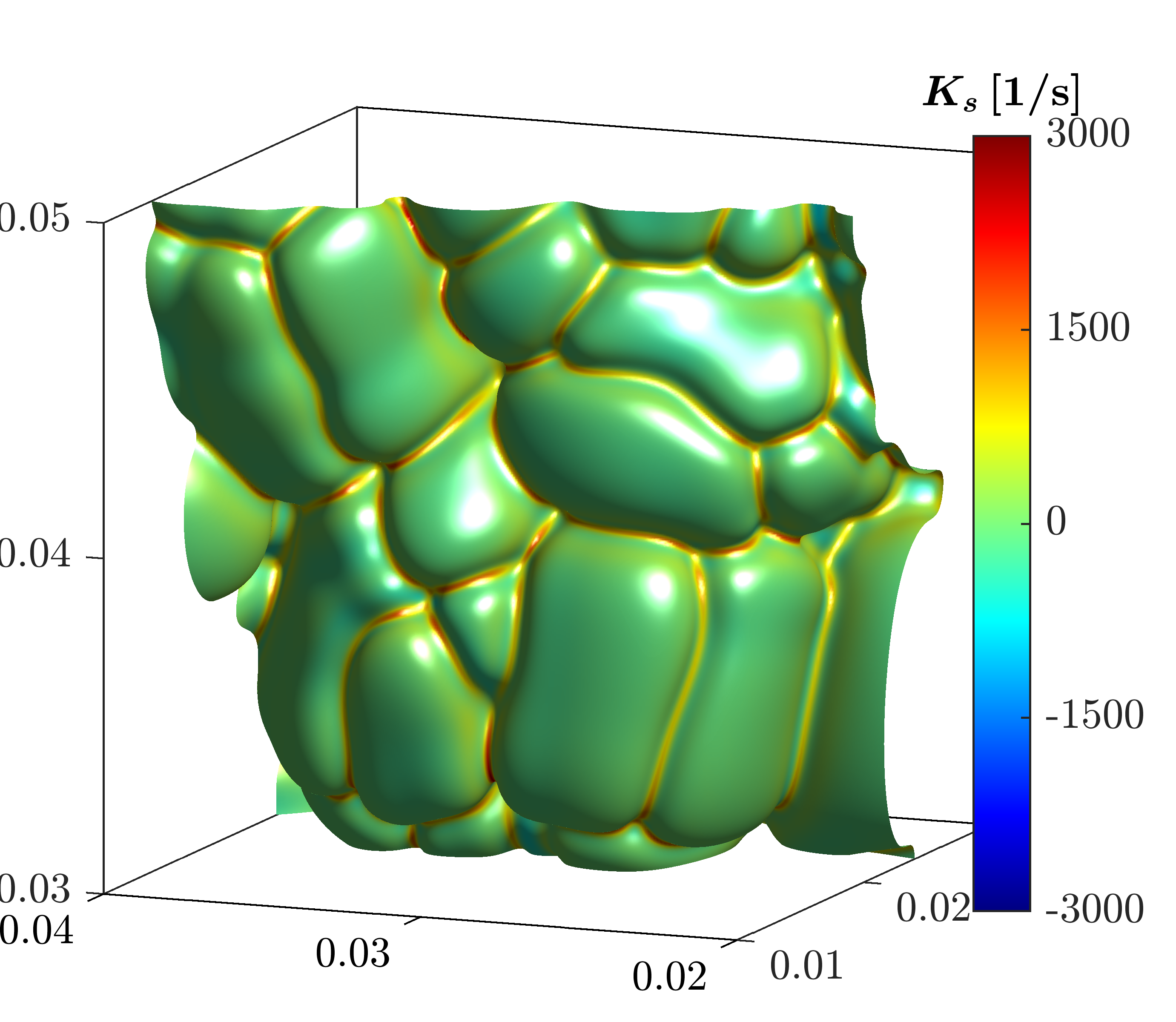}};
\end{minipage}
\begin{tikzpicture}[remember picture,overlay] 
	\node[text=black,scale=1] at (-5.6,0.1) (a) {$(b)$};
\end{tikzpicture}}
    \subfloat[]{\label{subfig:curvature}
    \begin{minipage}[b]{0.33\textwidth}
\tikz[remember picture] \node[inner sep=0pt,outer sep=0pt] (a) 
{\includegraphics[width=1\linewidth]{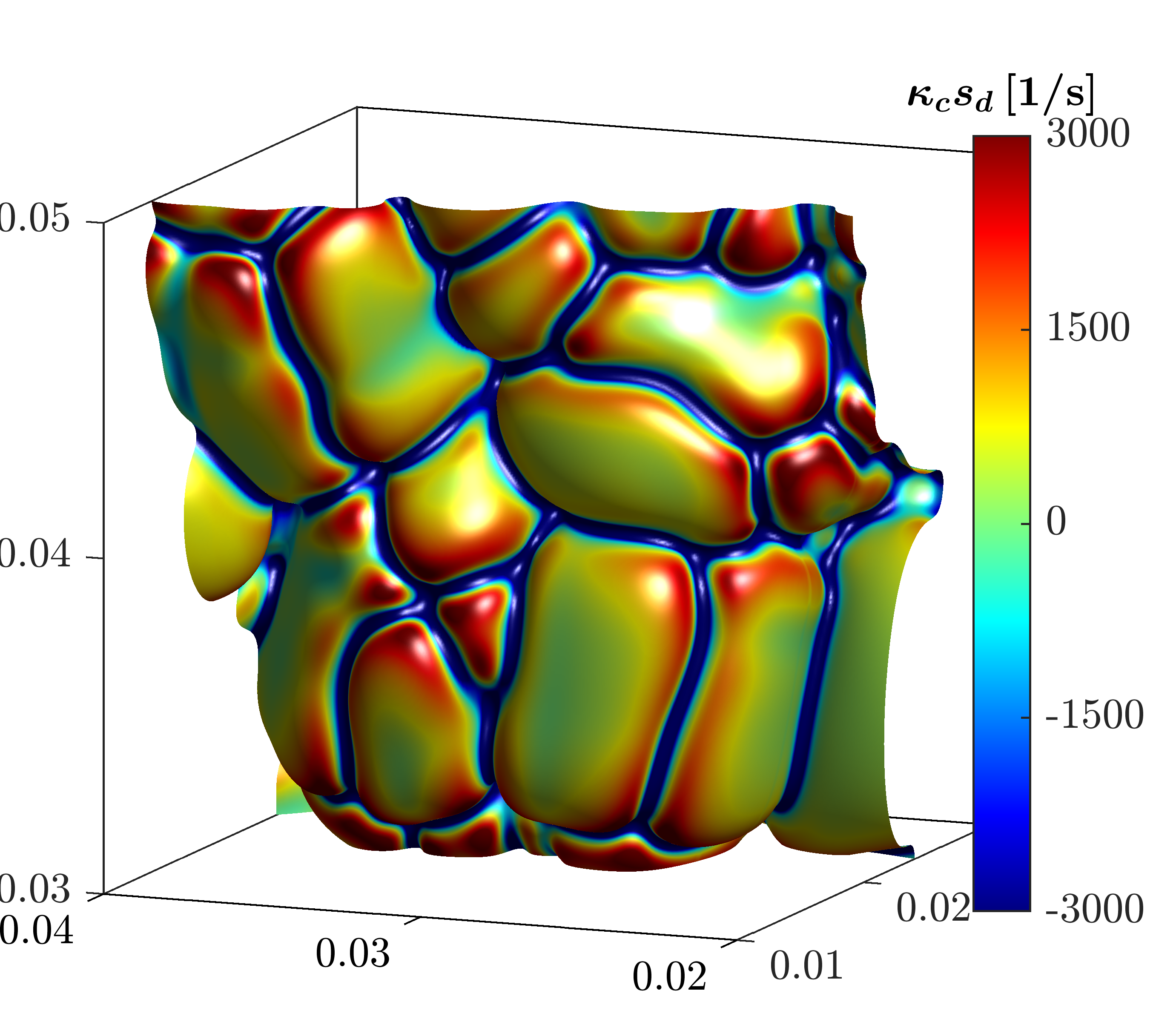}};
\end{minipage}
\begin{tikzpicture}[remember picture,overlay] 
	\node[text=black,scale=1] at (-5.6,0.1) (a) {$(c)$};
\end{tikzpicture}}
    \vspace{-6mm}
    \caption{\label{fig:KskappaK} Isosurface of $\psi = 0.8 $ colored by the local values of (a) stretch rate, (b) tangential strain rate, and (c) the curvature term for the 3D thermodiffusively unstable premixed hydrogen flame in a selected region to highlight the bulb-like flame structure. }
\end{figure}
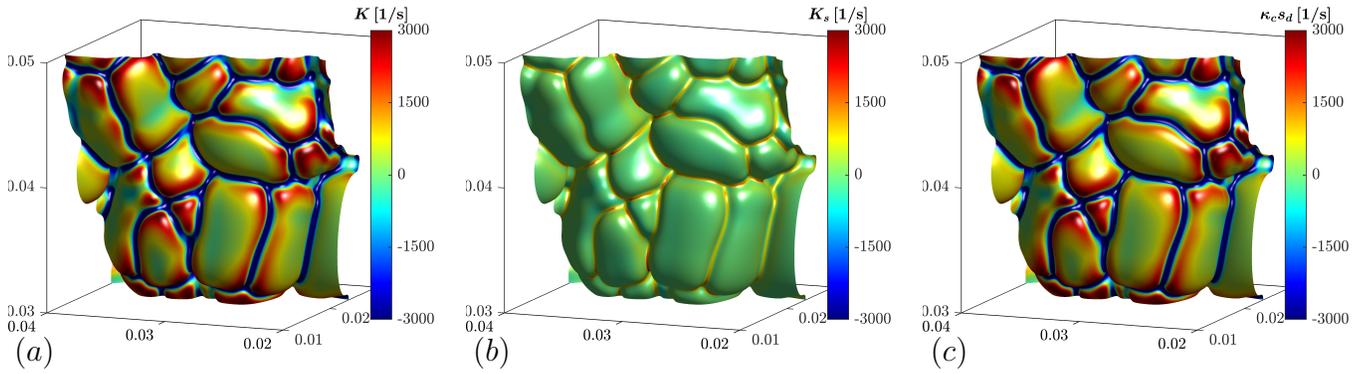

In this subsection, the mechanism for the formation and destruction of the distinct bulb-like flame structure is investigated. At first, the change of the flame surface area, which is determined by the stretch rate, cf.~Eq.~\eqref{eq:stretchRate}, is investigated. Figure \ref{fig:KskappaK} shows the isosurface of $\psi = 0.8 $ colored by the local values of stretch rate, tangential strain rate, and the curvature term for the 3D thermodiffusively unstable premixed flame in a selected region to highlight the bulb-like flame structure. The instantaneous distribution of the stretch rate $K$ in Fig.~\ref{subfig:stretch} shows that a strong formation of flame surface area ($K > 0$) takes place at the edges of the small positive-curvature cells, while the destruction of the flame surface area ($K < 0$) happens almost exclusively in the narrow negatively-curved regions. For the instantaneous distributions of the tangential strain rate $K_s$ and the curvature term $\kappa_c s_d$ in Figs.~\ref{subfig:strainRate} and \ref{subfig:curvature}, respectively, it can be observed that the tangential strain rate contributes to the formation of flame surface area in the negatively-curved regions, but otherwise, the curvature distribution dominates. In fact, the magnitude of the curvature term is more than twenty times larger than the tangential strain rate term.

\begin{figure}[!h]
    \centering
    \captionsetup[subfigure]{labelformat=empty}
    \subfloat[]{\label{subfig:KineRestor}
    \begin{minipage}[b]{0.42\textwidth}
\tikz[remember picture] \node[inner sep=0pt,outer sep=0pt] (a) 
{\includegraphics[width=1\linewidth]{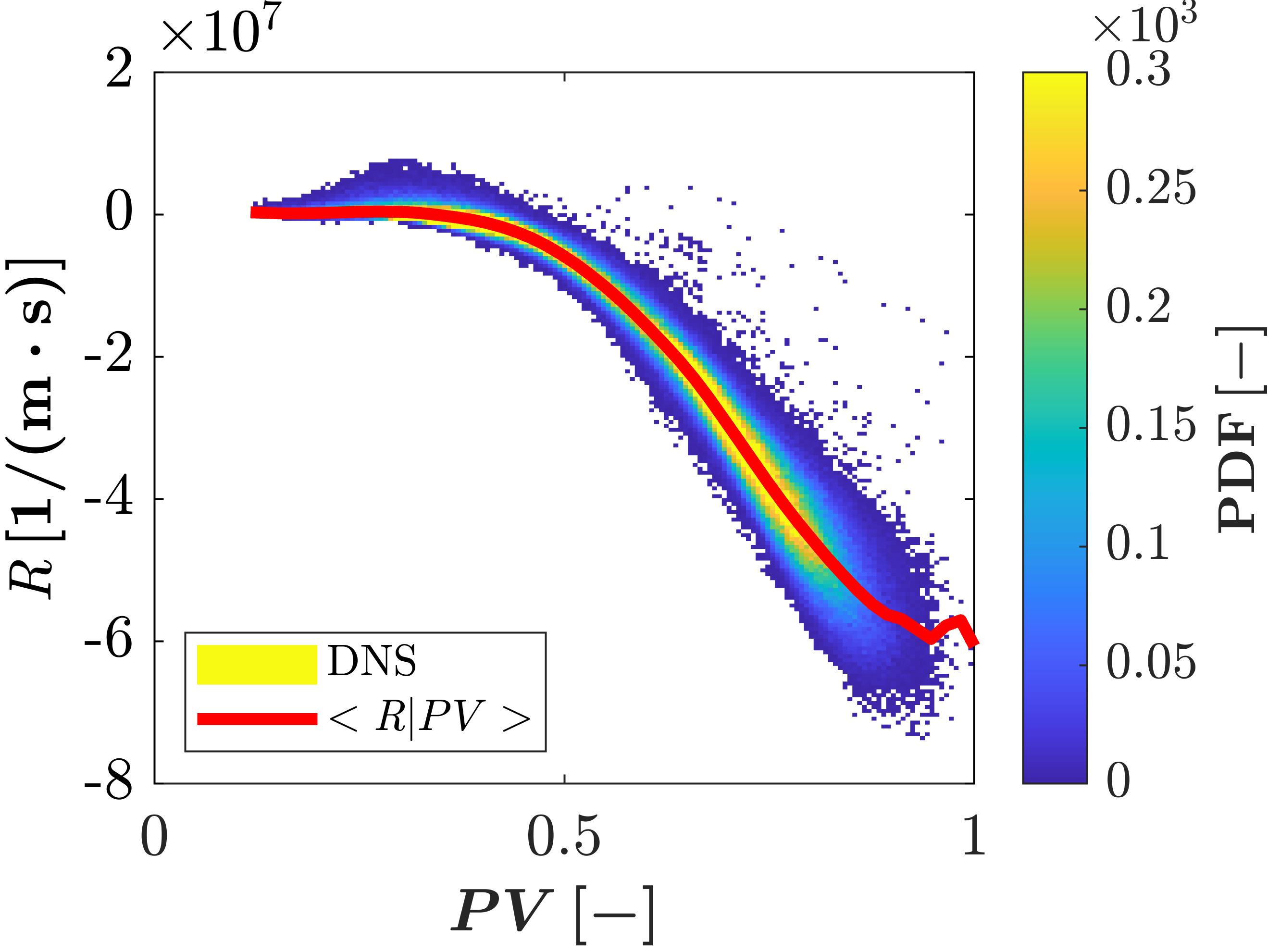}};
\end{minipage}
\begin{tikzpicture}[remember picture,overlay] 
	\node[text=black,scale=1] at (-6.6,0) (a) {$(a)$};
\end{tikzpicture}}
    \subfloat[]{\label{subfig:CurvDissi}
    \begin{minipage}[b]{0.42\textwidth}
\tikz[remember picture] \node[inner sep=0pt,outer sep=0pt] (a) 
{\includegraphics[width=1\linewidth]{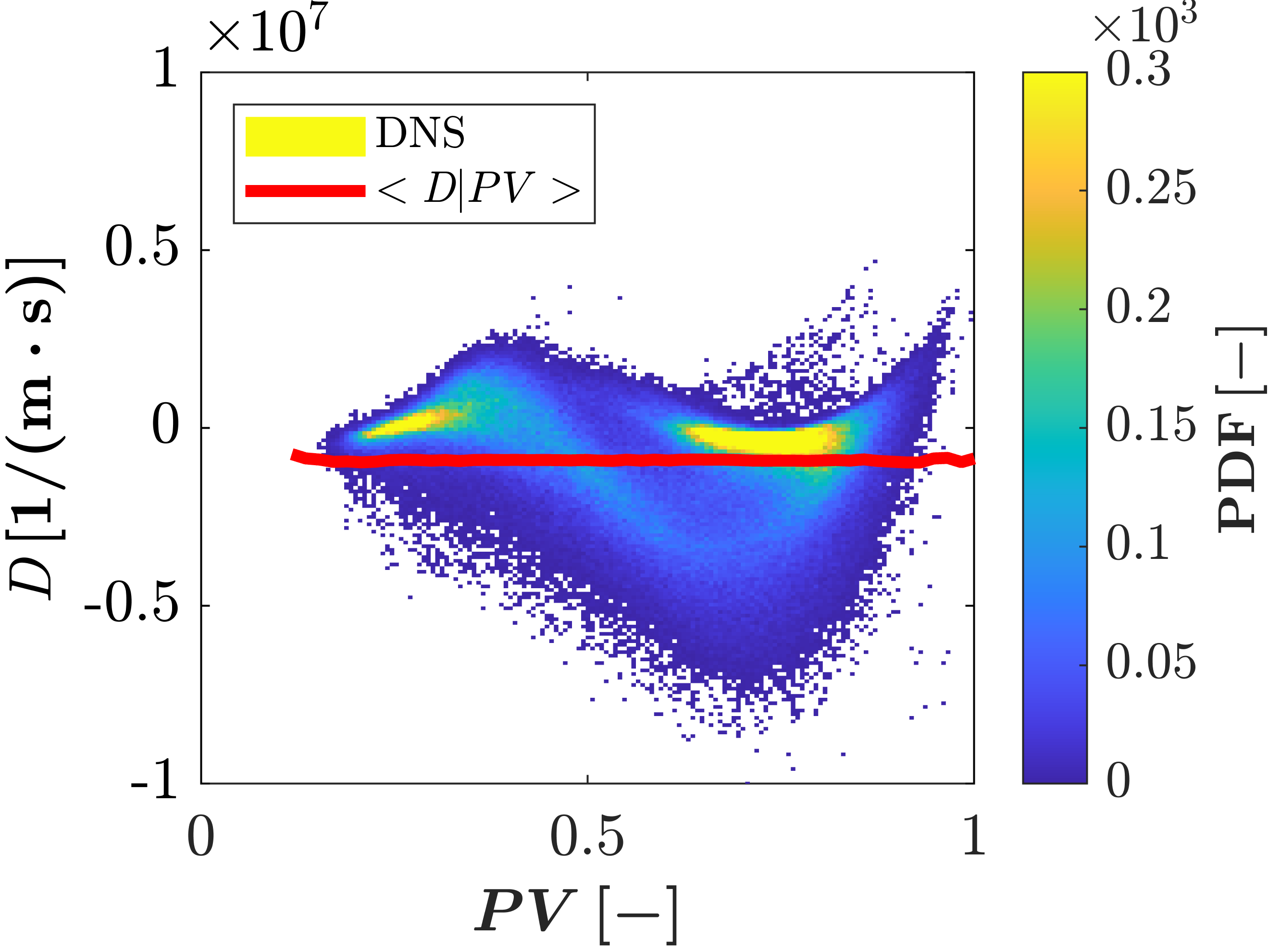}};
\end{minipage}
\begin{tikzpicture}[remember picture,overlay] 
	\node[text=black,scale=1] at (-6.5,0) (a) {$(b)$};
\end{tikzpicture}}
    \vspace{-6mm}
    \caption{\label{fig:KineCurv} Conditioned mean of the (a) kinematic restoration term, and the (b) curvature dissipation term contained in the balance equation for surface density function \citep{kollmann1998pocket}. The statistics are extracted from the whole domain within the reactive variable range of $ 0.79 < \psi < 0.81 $ around the flame front. }
\end{figure}
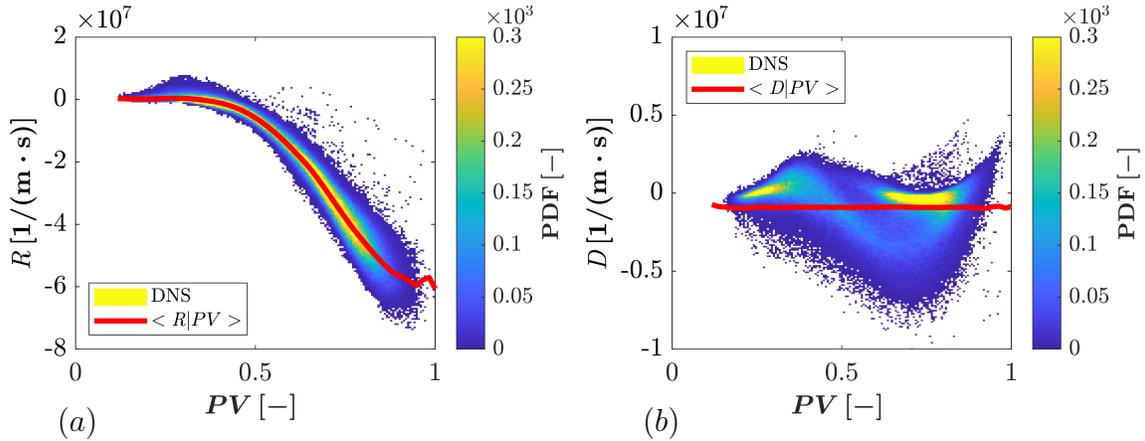

To quantify the contribution of kinematic restoration and curvature dissipation to the destruction of flame area, the balance equation for surface density function, $\sigma$, formulated by Kollmann and Chen \citep{kollmann1998pocket} is introduced. The kinematic restoration term contains two contributions from normal diffusion and reaction
\begin{equation}
\mathscr{R} = n_i \dfrac{\partial}{\partial x_i} \left( \dfrac{1}{\rho} n_j \dfrac{\partial \rho D_{\mathrm{H_2}} \sigma }{\partial x_j}  \right) + n_i \dfrac{\partial }{\partial x_i} \left(\dfrac{ \dot{\omega}_{\mathrm{H_2}} }{\rho} \right),
\end{equation}
while the curvature dissipation term can be written as
\begin{equation}
\mathscr{D} = n_i \dfrac{\partial \kappa_c D_{\mathrm{H_2}} \sigma }{\partial x_i},
\end{equation}
where $\dot{\omega}_{\mathrm{H_2}}$ and $D_{\mathrm{H_2}}$ are the production rate and mass diffusion coefficient of hydrogen, respectively. The quantitative values of the two terms shown in Fig.~\ref{fig:KineCurv} indicate that the magnitude of the kinematic restoration term is much larger than the curvature dissipation term. Thus, the destruction of flame area is dominant by the kinematic restoration rather than dissipation.

For the 3D simulations, the bulb-like structures appear to be formed and destroyed periodically, as also shown in the flame animation provided in the Supplementary Material. The mechanisms for this are very similar to those of the 2D flames discussed in detail by Berger et al.~\citep{berger2019characteristic}. However, there is one curiosity in the patterns of the 3D thermodiffusively unstable flame. As shown in Figs.~\ref{fig:3Dstructure} and \ref{fig:KskappaK}, the cells obviously assume rectangular shapes, which has not been explained. The formation of the rectangularly shaped flame front is due to the fact that corner-like structures  are strongly curved in two directions and hence have the highest curvatures, higher than the edges of the cells. As a result, they form the richest regions and propagate fastest with a component that is tangential to the overall flame surface, and as a result, they form stable rectangular structures. More detailed analyses for the formation and destruction of the bulb-like structures are provided in Sections 9 and 10 of the Supplementary Material.

\section{Summary and conclusions}
\label{Sec:5}

In this work, direct numerical simulations are conducted for thermodiffusively unstable laminar premixed hydrogen flames stabilized in 3D computational domains, in which NO$_\mathrm{x}$ formation is considered with detailed chemistry. The effects of confinement on the flame dynamics are quantified by increasing the domain size gradually until a constant burning velocity is obtained with the largest domain size, which is chosen according to the findings in ref.~\citep{berger2019characteristic}. The characteristic patterns of the thermodiffusively unstable hydrogen flame are investigated in detail, including the instantaneous flame structure, burning velocity, flame surface area, stretch factor, curvature distribution, and the cell size. In addition, the effects of computational setup (2D vs.~3D) on the distributions of the thermo-chemical quantities are quantified through conditional analyses in progress variable space and curvature space. Finally, the formation and destruction mechanisms of the distinct bulb-like flame structure are discussed. The main conclusions obtained from this work are summarized as follows:
\begin{enumerate}[(i)]
\item For the 3D thermodiffusively unstable premixed hydrogen flame studied, bulb-like structures with various length-scales can be observed in the flame front. Strong heat release enhancement is observed in positively-curved flame segments while quenching is seen in negatively-curved segments;
\item The global burning velocity and the flame surface area calculated with the domain with less than 50 flame thicknesses in the lateral direction are smaller than the largest domain with 100 flame thicknesses in the lateral direction, which indicates the existence of confinement effects in the 3D simulation. This is consistent with the findings for 2D studies \citep{berger2019characteristic, creta2020propagation}. The probability of large positive curvature values ($\kappa_c > 400 \, \mathrm{m^{-1}} $) in the smaller domains is higher than the largest domain, which leads to a higher probability of smaller cells;
\item Differences between 2D and 3D simulations are only seen in the range where the 3D simulations extend to substantially higher values in mixture fraction and progress variable. Such extended range causes the higher reactivity in the 3D simulations. This implies that the manifold observed in the 3D simulations can be well described from 2D simulations as long as high enough curvature values are present to form mixtures as rich as those in the 3D simulations;  
\item In the progress variable space, the conditioned mean mass fraction and production rate of major reactants and products do not show obvious differences between the 2D and 3D simulations, but the peak concentration/production rate of the highly diffusive species of H radical in the 3D simulation can be two/five times higher than the 2D simulation. The joint scalar/curvature PDFs are overall similar between 2D and 3D, but larger curvatures exist in 3D that correlate with larger mixture fraction values;
\item Both the formation and destruction of the flame surface area are governed by the curvature term, while the tangential strain rate plays a much less important role. The destruction of the flame surface is quantified due to the kinematic restoration with negligible contribution from the curvature dissipation. The small cells with higher propagating speed merge with the destructed large cells, which results in a periodic formation and destruction of the cellular flame front.
\end{enumerate}

\section*{Acknowledgments}

The authors acknowledge the generous support of the European Research Council (ERC) Advanced Grant (ID: 101054894). Xu Wen acknowledges support through Marie Sk\l{}odowska-Curie Individual Fellowship (ID: 101025581) awarded by the European Commission under H2020-EU.1.3.2 scheme. Part of the work was done during the 2022 CTR Summer Program at Stanford University. The financial support of the CTR and the fruitful discussions with Dr.~Kazuki Maeda and Dr.~Jonathan Wang from CTR, and Prof.~Thierry Poinsot from IMFT are greatly appreciated. Computational resources are provided by the Gauss Centre for Supercomputing e.V. on the GCS Supercomputer SuperMUC-NG at Leibniz Supercomputing Centre. We also thank the anonymous Reviewers for their valuable and constructive comments.








\bibliographystyle{model1-num-names.bst}
\bibliography{science}




\end{document}